\documentclass[usenatbib,referee]{mn2e} 
\usepackage{graphicx}
\usepackage{epsf}
\usepackage{paul}
\usepackage{lscape}
\usepackage{longtable}
\begin{document}
\title{Study of candidate Be stars in the Magellanic Clouds using NIR photometry and optical spectroscopy}
\author[Paul et al.]
{ Paul, K.T.$^{1}$,
Subramaniam, A.$^{2}$,
Mathew B.$^{2,3}$,
Mennickent R. E.$^{4}$,
Sabogal, B.$^{5}$
\thanks{email: paul.kt@christuniversity.in(PKT), purni@iiap.res.in(AS), blesson@prl.res.in(BM), rmennick@udec.cl(REM), bsabogal@uniandes.edu.co (BS)}\\
$^1$Dept. Physics, Christ University, Bangalore \\
$^2$Indian Institute of Astrophysics, II Block Koramangala,
Bangalore 560034, India \\
$^3$Astronomy \& Astrophysics division, Physical Research Laboratory, Ahmedabad, India \\
$^4$ Dept.of Astronomy, University of Concepcion, Concepcion, Chile \\
$^5$ Departamento de F\'{\i}sica, Universidad de los Andes,
      Cra. 1 No. 18A-10, Edificio Ip, Bogot\'a, Colombia\\
}
\date{\it Submitted: September 2011}
\pubyear{2011} 
\maketitle
\label{firstpage}
\begin{abstract}
\citet{mennickent2002} and \citet{sabogal2005} identified a large number of Classical Be (CBe) candidates ($\sim$ 3500) in the L\&SMC based on their photometric variability using the OGLEII database. They classified these stars into four different groups based on the appearance  of their variability. In order to refine and understand the nature of these large number of stars, we studied the infrared properties of the sample as well as the spectroscopic properties of a subsample. We cross-correlated the optical sample with the IRSF catalog to obtain the $J$, $H$, $K_s$ magnitudes of all the four types of stars ($\sim$ 2500) in the L\&SMC. Spectra of 120 stars belonging to the types 1, 2 and 3 were analysed to study their spectral properties.  Among the four types, the type 4 stars is the dominant group, with $\sim$ 60 and $\sim$ 65\% of the total sample in the LMC and the SMC respectively. The NIR colour-colour diagrams suggest that the type 4 stars in the LMC have a subclass, which is not found in our Galaxy or in the SMC. This subclass is $\sim$ 18\% of the type 4 sample. The main type 4 sample which is $\sim$ 49\% of the total sample has  NIR properties similar to the Galactic CBe stars and the SMC type 4 stars.  Though the new subclass of type 4 stars have high $E(B-V)$ $\sim$ 0.75, they are not located close to regions with high reddening. The type 3 stars ($\sim$ 6\% \& 7.3\% in the L\&SMC) are found to have large H$_\alpha$ $EW$ in the SMC and some are found to have large NIR excess. This small fraction of stars are unlikely to be CBe stars. 3 stars among the type 3 stars in the LMC are found to be Double Periodic Variables.  The type 2 stars are found in larger fraction in the SMC ($\sim$ 14.5\%), when compared to the LMC ($\sim$ 6\%). The spectroscopic and the NIR properties suggest that these could be CBe stars. The type 1 stars are relatively more in the LMC ($\sim$ 24\% ) when compared to the SMC ($\sim$ 13\%). The SMC type 1 stars have relatively large H$\alpha$ $EW$ and this class has properties similar to CBe stars. The spectroscopic
sample of type 1 stars which show H$_\alpha$ in emission and confirmed as CBe stars are more abundant
in the SMC by a factor of 2.6.  If the effect of metallicity is to cause more CBe stars in the SMC, when compared to the LMC, then type 1, type 2 and type 4 stars follow this rule, with an enhancement of 2.6, 2.4 and 1.3 respectively.

\end{abstract}

\section{Introduction}

Be stars are B-type stars with luminosity class III to V that show or have shown emission in Balmer 
lines like H$_\alpha$, which originate in a circumstellar disc. 
They occupy a region on or 
near the main sequence in the HR diagram, implying that they are still burning hydrogen in their core. 
These stars are rapid rotators and show both photometric and spectroscopic variability. 
Detailed studies of Be stars in environments with different metallicities, such as the Magellanic Clouds, 
have been performed in the recent past \citep{keller1999,dewit2003,mennickent2002,sabogal2005,martayan2010}.

\citet{mennickent2002} presented a catalogue of 1056 Be star candidates in the 
Small Magellanic Cloud (SMC) by studying light curve variations using OGLE II data base \citep{udalski1998,udalski2000}.
They classified these Be star candidates of the SMC in four categories: type 1 stars showing 
outbursts (139 stars); type 2 stars showing sudden luminosity jumps (154 stars); 
type 3 stars showing periodic or near periodic variations (78 stars); 
type 4 stars showing light curves similar to Galactic Be stars (658 stars). 
They also classified type 1 stars with luminosity jumps in their light curves as 
type1/type2 stars (18 stars). They suggested that type 4 could be Be stars. They also proposed
that some of the type 1 and type 2 stars might be Be stars with accreting white dwarfs in a
Be + WD binary, or they could be blue pre-main sequence stars showing accretion disc
thermal instabilities. Spectroscopy is needed to confirm the suggestion that
some of these stars are Be stars. Also, more studies, especially in the near-infrared (NIR) are
required to confirm the pre-main sequence hypothesis.
On the other hand, they suggested that type 3 stars should not be
linked to the Be star phenomenon at all. 

Based on a similar inspection of OGLE II data, \citet{sabogal2005} classified Be 
candidates in the Large Magellanic Cloud (LMC) also as type 1 (581 stars), type 2 (150 stars), 
type 3 (149 stars), type 4 stars (1468 stars) and type 1/type 2 stars (98 stars). 
However many of the type 4 stars in the LMC are found to be reddened 
and located parallel to the main sequence, this feature was not found in the same diagrams 
of the SMC. The photometric properties of type 1 and type 3 stars on the LMC are very different
from those of the SMC. Thus, the various types of stars identified based on variability
seem to differ between the LMC and the SMC. 

The above studies identified a large number of candidate Be stars which can be used 
to derive the parameters  that are responsible for the Be-phenomenon as these types of stars in the MCs are metal poor when compared  to the Galactic Be stars. The above classification was based only on photometric variability. 
It is desirable to get some more properties like their NIR magnitudes and 
colours to understand these stars.
The aim of this paper is to study the NIR properties of various types of Be star candidates in  the L\&SMC 
by cross-matching IRSF (NIR) and OGLE II (optical) catalogs. We also present results from a  spectroscopic study of  70 stars from types 1, 2 and
 3 in the SMC 
and 49 stars belonging to type 1 and type 3 in the LMC. The spectral features are used to identify their
spectral class and to see whether they show some properties of Be stars. The paper is arranged as follows.
Details of the NIR photometric data as well as the spectroscopic data are presented in section 2. The
results of the cross-correlation between the optical and NIR properties as well the spectral classification
are presented in sections 3 and 4. The results are presented in section 5. Discussion and conclusion are presented in sections 6 and 7 respectively.

\section{Photometric data and Spectroscopic observations}

Catalogs of Be star candidates in the SMC and the LMC as identified by \citep{mennickent2002}
($\sim$ 1000 stars) and
\citep{sabogal2005} (2446 stars) respectively were  used. In order to obtain their NIR properties, 
we used the  
NIR IRSF catalog (\citep{kato2007}; http://pasj.asj.or.jp/v59/n3/590315) which has $JHK_s$ photometric data for about 15 million point 
sources spread over a 40 deg$^2$ area of the LMC and 2.7 million sources spread over a 11 deg$^2$ area of the SMC. The 10$\sigma$ limiting magnitudes of the above catalog are 18.8, 17.8 and 16.6 mag at $J$, $H$ and $K_s$ respectively. This catalog has better spatial resolution than the 2MASS catalog.
The optically identified stars were cross-matched with NIR IRSF catalog to confirm its candidature in IRSF. 
Matches between both catalogs were found by comparing the RA and DEC co-ordinates in both the catalogs. 
We considered the closest candidates as NIR counterpart, with an upper limit of 0.001 degree (3.6 arcsec) for separation.
We could cross-identify the NIR counterpart of 1640 stars among the 2348 candidate Be stars in the LMC (excluding the type 1/2 candidates) and 839 stars from the sample of 1029 stars in the SMC. Hence the cross identification is 70\% successful in the LMC and 80\% in the SMC. Due to crowding we could not cross-identify rest of the stars. These stars were used to study the NIR properties of the various types.

Optical spectra were obtained for 120 candidate stars in the LMC and the SMC. Spectroscopic observations were conducted at the Cerro Tololo Inter-American Observatory (CTIO) and the Las Campanas Observatory (LCO) during 4 nights of October 2002 and 4 nights of November 2003, respectively. At the CTIO we used the 1.5-m telescope with the Cassegrain Spectrograph and the Loral 1-K detector. This spectrograph details are available in the web page:
http://www.ctio.noao.edu/spectrographs/60spec/60spec.html.
 We obtained 57 blue spectra with the grating $n_o$. 26 tilted at 16.2 degree and a slit width of 1.5 arcseconds, yielding a wavelength range of 3700$-$5500 \AA{}. 52 red spectra were obtained with grating $n_o$. 36 tilted 28.7 deg,  resulting in a wavelength range of 5700$-$7000 \AA{}. In both cases the resolution was about 3.7 \AA{}. The spectra were reduced with standard IRAF routines, like apall and doslit, including bad pixel rejection, spectroscopic flat fielding, 1d extraction and wavelength calibration. The typical error in the wavelength calibration is 5 km/s in H$\alpha$.

In addition, 48 spectra were obtained in LCO with the Modular Spectrograph and the SiTe2 detector.  Technical details for this spectrograph can be found in
the web page of the LCO observatory\footnote{http://www.lco.cl}.  The combination of grating $n_o$. 600 blazed at 5000 \AA{}, with  a slit width of 1.5 arcseconds yielded a spectral range of 3870$-$6100 \AA{}, and a resolution of 2.5 \AA{}. The log of observations are given in tables 1 and 2 (see appendix). The spectra were reduced and wavelength calibrated using the standard IRAF tasks. The typical error in the wavelength calibration is 12 km/s in H$\beta$. These spectra were used to identify the spectral lines, determine spectral types and measure equivalent width.

\section{Analysis of NIR colours}
We cross-correlated 1640 stars (70\%) identified by \citet{sabogal2005} with the IRSF data in the LMC. 
Among these, 399 are type 1 stars, 92 are type 2 stars, 91 are type 3 
stars and 989 stars are type 4 stars. 
The SMC sample contains 839 stars, of which 89 are type 1 stars, 131 are type 2 
stars, 65 are type 3 stars and 554 are type 4 stars.

Adopting an average value of $E(B-V)$ = 0.1 and $R=A_v/E(B-V) = 3.1$, on the basis of previously published values, colour magnitude diagrams (CMDs) 
with $(B-V)_0$ vs $V_0$ were plotted. Figure 1 and Figure 2 show 
$V_0$ versus $(B-V)_0$ CMDs for the cross matched stars (type 1 to type 4 shown in different colours)
 for the LMC and the SMC respectively. These figures show the location of all the types in the 
CMD, which indicates that the cross-identified stars span the
full range of locii similar to the optically identified sample (in comparison with the figures given in \citet{sabogal2005} and \citet{mennickent2002}). 
We notice the red sequence of type 4 stars located parallel to the main sequence in the 
range of $(B-V)$ = 0.4 to 0.7 mag, in Figure 1 (LMC), which is absent in Figure 2 (SMC). This feature
was noticed by \citet{sabogal2005}. This shows that the cross-correlated sample contains these 
peculiar stars also. In order to quantify and thus verify whether we have evenly sampled all the types in both the galaxies, we estimated the fractions of various types.

\renewcommand{\thetable}{3}
\begin{table*}
\begin{center}
\caption{Number of various types of stars identified in optical and infrared in the LMC and the SMC. Relative fraction of various types are indicated in the parenthesis. Columns 6 and 7 represent the ratio of various types of stars in the LMC to the SMC}
\end{center}
\begin{tabular}{|l|l|l|l|l|l|l|}
\hline
\hline
&\multicolumn{2}{c|}{$LMC$}&\multicolumn{2}{c|}{$SMC$}&\multicolumn{2}{c|}{$LMC/SMC$}\\
\cline{2-7}
Type & Optical &  Optical + IR & Optical & Optical + IR & Optical & Optical + IR\\
\hline
1&581(0.24)&399(0.24)&139(0.13)&89(0.10)&4.17&4.48\\
2&150(0.06)&92(0.056)&154(0.15)&131(0.156)&0.97&0.7\\
3&149(0.06)&91(0.055)&78(0.07)&65(0.077)&1.9&1.4\\
4&1468(0.60)&989(0.60)&685(0.65)&554(0.66)&2.14&1.76\\
1/2&98(0.04)&69(0.042)&--&--&--&--\\
Total&2446&1640&1056&839&2.3&1.95\\
\hline

\end{tabular}
\end{table*}
The fraction of the four types of stars in the original sample and in the cross-correlated sample are given in Table 3. The first column indicates the types, followed by the number of stars identified in the
optical and then the cross-correlated sample for both the galaxies. The last two columns give the ratio of stars in the two galaxies (ratio = $LMC/SMC$). The fractions of optically identified type 1 to type 4 stars are similar to the fraction in the cross-correlated sample. This suggests that the cross-identification has sampled all the types similarly and this subset can be used
to study the properties of various types in the NIR. Later in the discussion we will use this Table to compare the relative population of the four types in the L\&SMC. We compared the magnitude distribution of stars identified in the NIR with
the original optical sample, in order to estimate how much stars are missed due
to the limiting magnitude of the NIR photometry. Since the SMC is more distant
compared to the LMC, there is a possibility that relatively more fainter stars are missed
out in the SMC. We compared the distribution in both the galaxies and we find
that the number of stars missed in the fainter limit is similar in both the
galaxies. Hence the NIR+optical sample discussed here is not biased to brighter stars in the SMC.

\begin{figure}
\epsfxsize=10truecm
\centerline{\epsffile{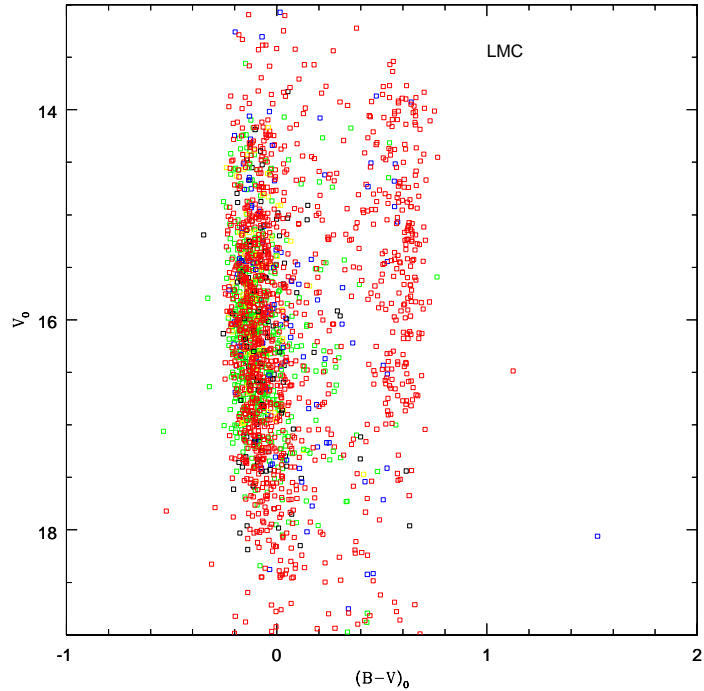}}
\caption{$V_0$ versus $(B-V)_0$ diagram of the total sample (type 1 in green colour, type 2 in black, type 3 in blue and type 4 in red colour) of the cross-correlated stars in the LMC.
}
\end{figure}
\begin{figure}
\epsfxsize=10truecm
\centerline{\epsffile{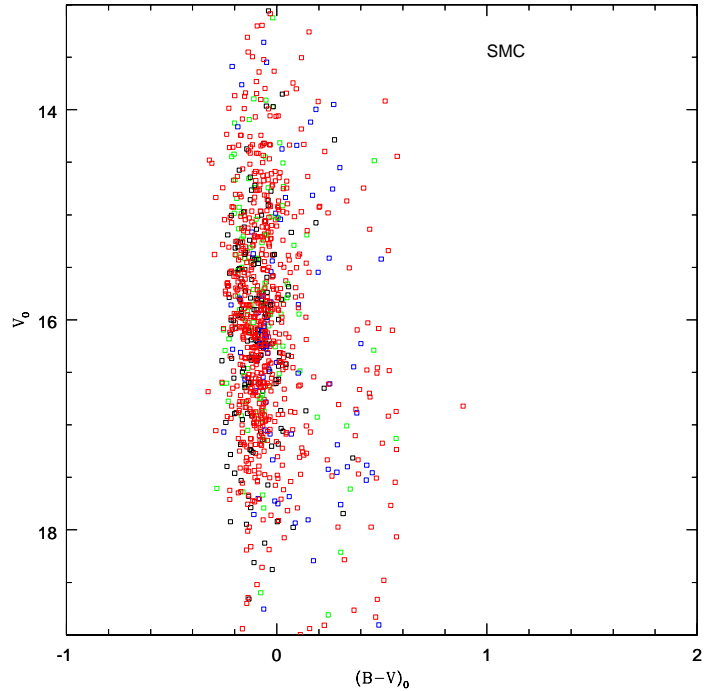}}
\caption{$V_0$ versus $(B-V)_0$ diagram of the total sample (type 1 in green colour, type 2 in black, type 3 in blue and type 4 in red colour) of the cross-correlated stars in the SMC.
}
\end{figure}
We estimated the de-reddened $(J-H)_0$ and $(H-K)_0$ values of the cross matched 
sources (1640 stars in the LMC and 839 stars in the SMC) using the IRSF data. 
We assumed an average reddening of $E(B-V)$ = 0.1 mag, which is basically the median of reddening
estimated by \citet{indu2011} using the bright main-sequence stars. Though the bar region of the LMC does not show
large reddening, some of the individual stars could show reddening different than the above value.
Since we do not derive any quantities for various types, the above assumption is justified. 
Reddening corrections were made using the following formula 
\citep{bessell1988} by taking $E(B-V)$ = 0.1 mag.
$$(J-H)_0 = (J-H)-E(J-H)$$
$$(H-K)_0 = (H-K) -E(H-K)$$
$$E(J-H) = 0.37*E(B-V )$$
$$E(H-K)=0.19*E(B-V)$$
Colour-Colour diagrams (CCDm), were plotted with $(H-K)_0$ Vs $(J-H)_0$ for all the types (type 1, 2, 3 and 4) of stars. 
Figures 3 and 4 show CCDm for the LMC and the SMC respectively for all types. 
In order to identify the location of the stars we over 
plotted the main-sequence, the reddening vectors and the region of T-Tauri stars. 
The location of T-Tauri stars is shown as the dashed straight line \citep{meyer1997}. 
The location of Be stars is taken from \citet{dougherty1994} and the location of Herbig Ae/Be stars is 
taken from \citet{hernaandez2005}.
\begin{figure}
\epsfxsize=10truecm
\centerline{\epsffile{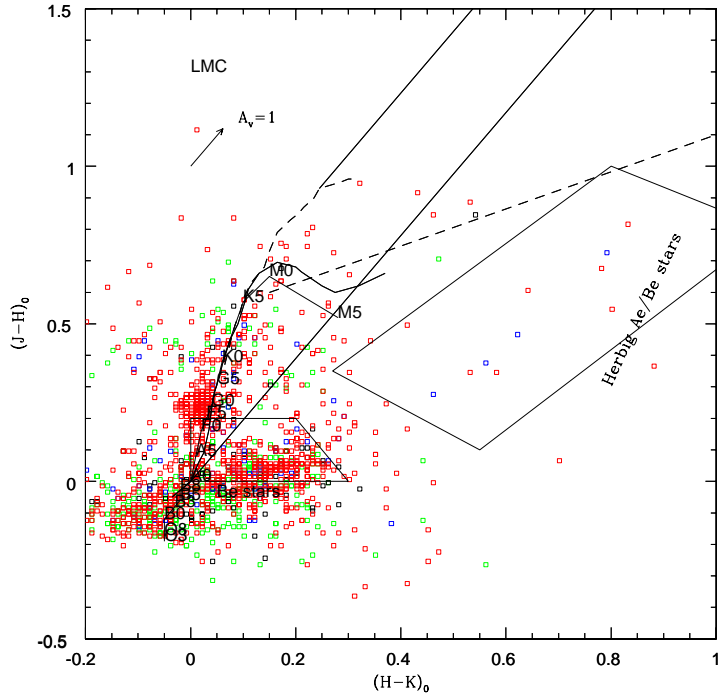}}
\caption{ NIR colour-colour diagram of the  Be candidates in the  LMC (same colour code as in Figure 1). The location of MS and the T-Tauri stars and reddening vectors are also shown. Reddening vector indicates $A_v = 1$
}
\end{figure}
\begin{figure}
\epsfxsize=10truecm
\centerline{\epsffile{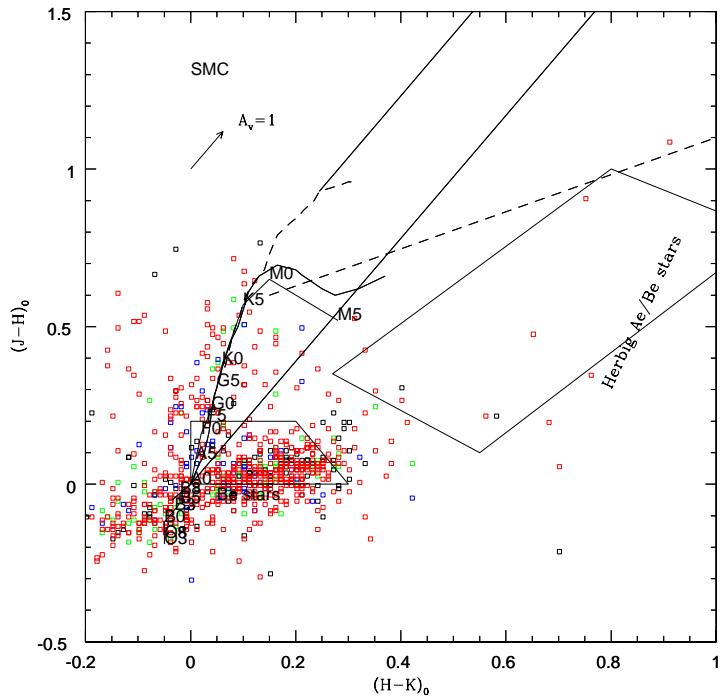}}
\caption{
NIR colour-colour diagram of the Be candidates in the SMC (same colour code as in Figure 1), rest same as Figure 3.
}
\end{figure}
The figure for the LMC (Figure 3) shows that most of the stars are populated near the $(J-H)_0$ $\sim$ 0.0, 
with a small range in
$(H-K)_0$, in a more or less horizontal band. We do also notice that some stars occupy a location above this band, on the MS, giving rise to a clumpy appearance. The stars in this clump are mostly type 4 stars. Apart from these two prominent features, some stars are found to be scattered in the diagram. The SMC figure (Figure 4) also shows the  horizontal band like distribution, but the clumpy population is not found in the SMC. Some amount of scatter is also noticed similar to the LMC. 
Thus, the NIR analysis suggests a different sub population among the type 4 stars in the LMC,
which is not found in the SMC. We shall explore these stars further in the next section.
We identify some sources with large NIR excess in the LMC, belonging to type 3 and type 4.
The SMC has fewer sources with NIR excess. 
 The distribution of the four types stars along the NIR colours, $(J-H)_0$ and $(H-K)_0$, are shown for the L\&SMC in figure 5. We compare the four types between the two galaxies using these histograms. The histograms not only show the distribution, but also compares the relative number of stars in various types in the two galaxies. We discuss the properties of various types and compare them between the LMC and the SMC in the following sections.

\section{Spectroscopic analysis}
Our spectroscopic sample contains 44 spectra in the SMC (17 type 1, 12 type 2 and 15 type 3) and 
13  spectra in the LMC (1 type 1, 12 type 3) in the blue region and 42 spectra in the SMC (15 type 1, 8 type 2 and  19 type 3) and 20 spectra in the LMC (8 type 1, 3 type 2 and 9 type 3) in the red region.
The reason for studying spectroscopically only type 1, type 2 and type 3, omitting type 4 stars at this stage, was that the type 4 stars were very likely Be stars. On the other hand, due to their atypical photometric variability, the nature of type 1, type 2 and type 3 star was not so obvious.  A preliminary study of spectra of type 4 stars in the SMC obtained with the ESO UVES spectrograph was presented in \citet{sabogal2011}, where the spectra showed H$_\alpha$ lines in emission. The detailed paper 
\citep{sabogal2012}   
based on the UVES spectra of type 4 LMC/SMC Be stars is in preparation. These two studies suggest that the type 4 stars likely to be CBe stars. 

The equivalent width  of the spectral lines were estimated using routines in IRAF. 
For spectroscopic classification, we compared the spectra of the Be candidate stars with those 
of the standard stars from \citet{jacoby1984} in the interval 3700$-$7000\AA{}. The classification was done by comparing the intensity of HI and HeI absorption lines of the candidate star in 3800$-$4500\AA{} region with the
library spectra.  The lines used for classification are HI 3835, 3970, 4101, 4340\AA{} and HeI 4026, 4121, 4143, 4388, 4471\AA{}. It is known that the spectral lines will be affected by emission component. Hence we have preferred the blue end of the spectrum which is least affected by contamination. Our idea was to derive a broad picture
about the spectral type rather than being accurate about the subclass. The derived spectral types are listed in Tables 4 to 9. Figures 13 and 14 show the representative spectra of Be candidates in the blue and red region from the sample. We are only able to estimate a range of spectral type. The luminosity classification is less accurate such that, we could not classify them as V or III. For B type stars, we estimated the spectral types from the equivalent width of the He{\sc i} lines assuming that our sample is composed of dwarf stars. For most of the spectroscopic sample H$\alpha$ is detected in emission. In the case of type 3 stars in the SMC, \citet{mennickent2006} have classified these stars using spectra of better resolution and we have shown both the classification. It can be seen that in most of the cases, the spectral types are very similar and this validates our method of classification. We have estimated radial velocity for most of the stars with spectra. The estimated radial velocities were corrected for heliocentric velocities. These estimated velocities are also tabulated in Tables 4  - 9 (except Table 7), along with the error in the velocity and the number of lines used for estimation. Wherever any previous measurement is available, they are also compared and discussed. These radial velocity measurements can be used to trace whether they are variables, which could point to their binary nature. 

For stars with spectra, $(J-H)$ and $(H-K)$ colours were de-reddened using the $E(B-V)$ values estimated based on their locations in the LMC and the SMC. The reddening towards these stars were taken from a reddening map estimated using the main-sequence stars \citep[map-B, figure 3]{indu2011}. These authors estimated an average reddening map for stars located near the
turn-off stars in that region. This reddening map is ideal for early type stars located on the MS. Since it is known that the reddening in the LMC depends on the tracer used, reddening map appropriate for the population studied is used.

\section{Results}
Our results of the NIR analysis and the spectral analysis are presented for the four types of stars. We do not have spectra for type 4 stars, and hence the results and conclusions for this class are based only of the NIR and optical photometric properties.

\subsection{Type 4 stars}
The type 4 stars are found to be the largest among the sample in both the Clouds. Their fractions
are 60\% \& 65\% in the LMC and the SMC respectively, suggesting that the fractions are similar in these two galaxies. Thus more than half of the total sample fall in this group, making it the single most important type among the four types studied.  
\citet{mennickent2002} suggested that the type 4 stars are likely to be Classical Be stars, 
since they show rather irregular photometric variation similar to the Galactic
counterparts. Therefore, in order to confirm the above, we compared the NIR properties of the Galactic Be stars 
along with the LMC and the SMC type 4 stars. We used data of the 
Galactic Be star candidates from \citet{mathew2008}. The NIR photometric magnitudes for all these candidate 
stars are taken from 2MASS data base. The $(J-H)$ and $(H-K)$ colours obtained were transformed to 
\citet{koornneef1983} system using 
the transformation equations by \citet{carpenter2001}. The colours were de-reddened. The $(J-H)_0$ and $(H-K)_0$ colours 
obtained for the Galactic Be stars were overplotted in the CCDm of the LMC and the SMC 
type 4 stars (Figure 6). 
It can be seen that the Galactic Be stars occupy the region which coincides with a horizontal band. Thus, the
type 4 stars which are located in this region are similar to the Galactic Be stars, with respect to their NIR colours.
A clumpy distribution of type 4 stars located above the  band like distribution is found in the LMC. This group does not have any counterpart in our Galaxy, or in the SMC.  The histograms shown in 
figure 5 suggest that more type 4 stars in the LMC have redder $(J-H)_0$ colours, compared to the SMC and this is due to the population which appears as a clump in the NIR CCDm. The distribution along $(H-K)_0$ colour suggests that the LMC type 4 has a bluer component within the range observed and this component is
the contribution from the new subgroup which appears as a clump. The colour range populated by the
type 4 stars are similar in both the galaxies.
Thus, we can summarise that type 4 stars in the LMC fall in two
groups, one with NIR properties similar to those of the Galactic Be stars, and a new sub-class with different NIR properties. This new sub-group does not have any NIR excess. Their location in the NIR diagram suggests that these may 
be A-F type stars located on the MS, or highly reddened OB stars. The type 4 stars in the SMC have NIR properties similar to the Galactic Be stars.  \citet{bonanos2010} identified a similar red
sequence among Be stars in the SMC, derived from a very different sample. They found these stars to be redder than the MS stars by about 0.7 mag. It is not clear whether these stars are similar to those
we have identified in the LMC.

It will be interesting to find the location of stars in the new sub-group in the optical CMD. Figure 7 shows the $V_0$ vs $(B-V)_0$ CMD of type 4 stars which appear as a clump in the NIR CCDm. These stars occupy the reddened parallel sequence in the
optical CMD. Among the 1468 type 4 stars, about 265 stars ($\sim 18\%$) show this property. In the cross-correlated sample, we find that among the 989 stars, 216 stars (21.8\%) appears as a separate group. Thus the cross-correlated sample also has similar fraction of the new subgroup. In the LMC, nearly 22\% of the type 4 stars appear as a separate group and hence only 47\% of the cross-correlated sample are type 4 stars having properties similar to Galactic Be stars. On the other hand, the SMC has about 65\% of the cross-correlated type 4 stars which are similar to Galactic Be stars. The fraction of type 4 stars, similar to the Galactic Be stars is relatively more abundant in the SMC. The enhancement is found to be 1.3 times in the SMC, with respect to the LMC, as estimated from the optically identified sample. Since the Be phenomenon is enhanced in low metallicity environment of the SMC, when compared to the LMC or our Galaxy as found by \citet{martayan2006,martayan2007}, the larger fraction of type 4 stars found in the SMC falls in line with the above argument.

It will be interesting to decipher the nature of the new subgroup. One possibility is that this new group of stars could be highly reddened OB stars. We checked whether
the reddening in the optical CMD required to create the parallel sequence is consistent with the reddening required 
in the NIR to create the clump like distribution. We estimated that the parallel sequence has a 
reddening of $E(B-V)\sim$ 0.75 mag more than the MS counterparts. This corresponds to $E(J-H) \sim$ 0.266 and $E(H-K) \sim$ 0.137 mag.
A plot showing the unreddened parallel sequence in the optical and NIR is shown in Figure 8,  assuming the above reddening values. 
It can be seen that the estimated reddening in the optical bands more or less fits the NIR reddening also, since the
unreddened sequence falls almost along the horizontal sequence of likely Be stars.
In the optical CMD, after correcting for the extinction, the parallel sequence is
found to occupy the brighter part of the MS. This as well as the unreddened location in the NIR diagram suggests
that these stars could be more massive than the normal type 4 stars.
We estimated the average reddening of the location in which these stars are located by comparing their locations in the extinction map estimated by \citet{ Harris2004} which provide reddening to individual stars. Figure 9 shows the histogram of the Av around the reddened type 4 stars. The average value we obtained is around 0.5 mag, suggesting that the $E(B-V)$ around these stars are less than 0.2 mag. The reddening required to make these stars fall back on the MS is 0.75 mag. For the $E(B-V)$ to be 0.75, Av should be more than 2.0, and the locations do not have such large extinction. Thus these stars are not likely to be associated with regions of high reddening. On the other hand, these stars could be individually highly reddened due to large mass loss, in which case it is highly unlikely that all the stars should have the same high reddening, instead of  a range in reddening. In either case, these stars are highly interesting and might turn out to be a new class of objects. Thus, it is necessary to obtain spectra of normal and the new group of type 4 stars to understand their nature. 
\begin{figure}
\epsfxsize=10truecm
\centerline{\epsffile{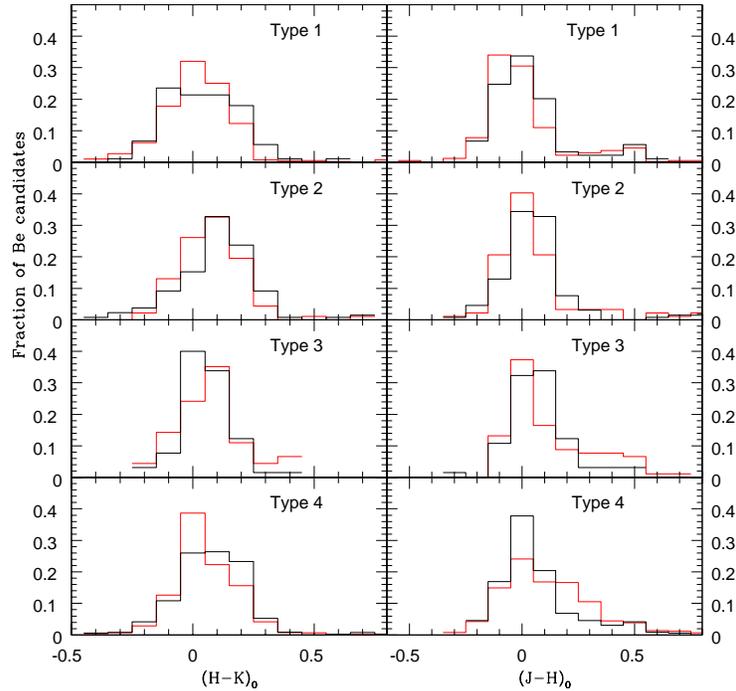}}
\caption{ Histograms of $(H-K)_0$ and $(J-H)_0$ colors of 1640 Be candidates  in the LMC (shown in red color) and 839 Be candidates  in the SMC (shown in black color) divided into various types (type1 to type 4).
}
\end{figure}

\begin{figure}
\epsfxsize=10truecm
\centerline{\epsffile{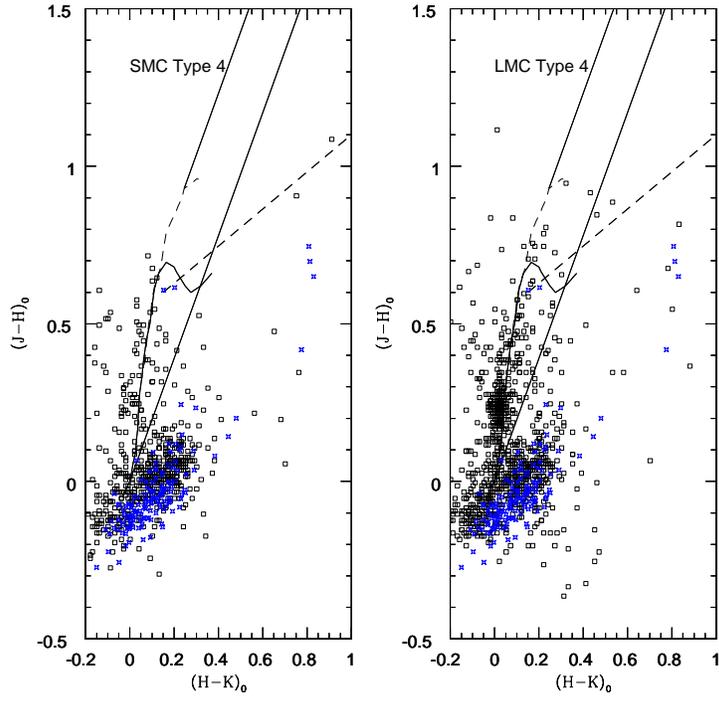}}
\caption{ NIR colour-colour diagram of the type 4 Be candidates in the  SMC and the LMC. The Galactic Be stars are shown as blue points.
}
\end{figure}
\begin{figure}
\epsfxsize=10truecm
\centerline{\epsffile{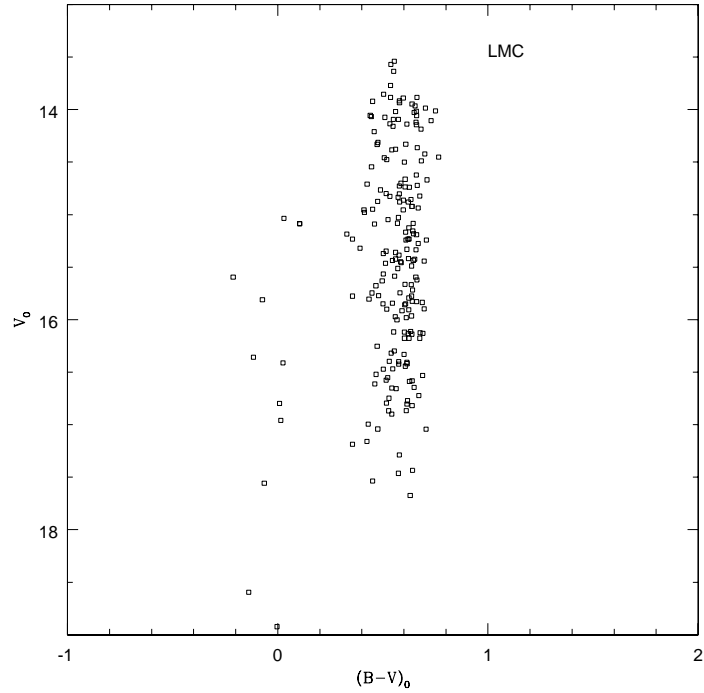}}
\caption{ The $V_0$ versus $(B-V)_0$ CMD of type 4 stars which appear as a clump in the NIR CCDm.
}

\end{figure}

\begin{figure}
\epsfxsize=10truecm
\centerline{\epsffile{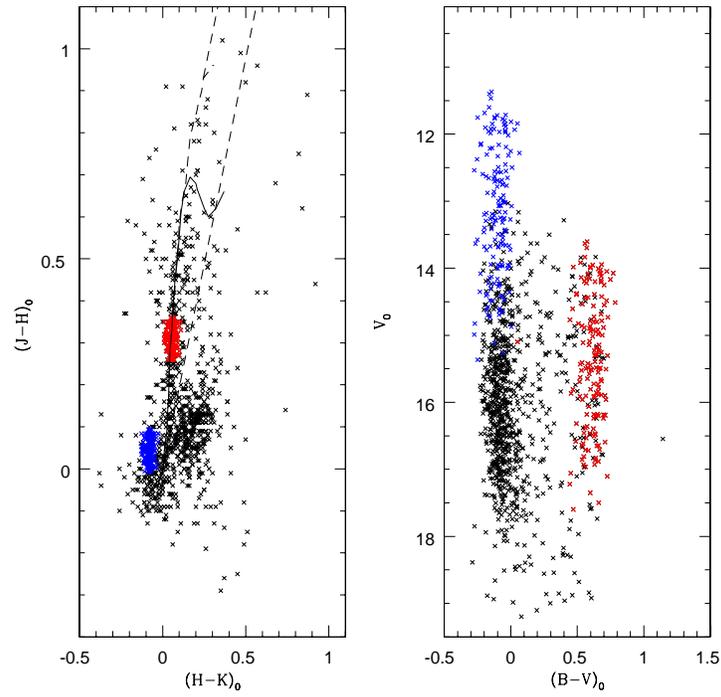}}
\caption{ CMD and CCDm showing all types of stars. The type 4 stars in reddened parallel sequence are shown in red colour along with all other types (shown in black colour). If the type 4 stars forming the parallel sequence is de-reddened, it will occupy the position shown with blue points 
}
\end{figure}
\begin{figure}
\epsfxsize=10truecm
\centerline{\epsffile{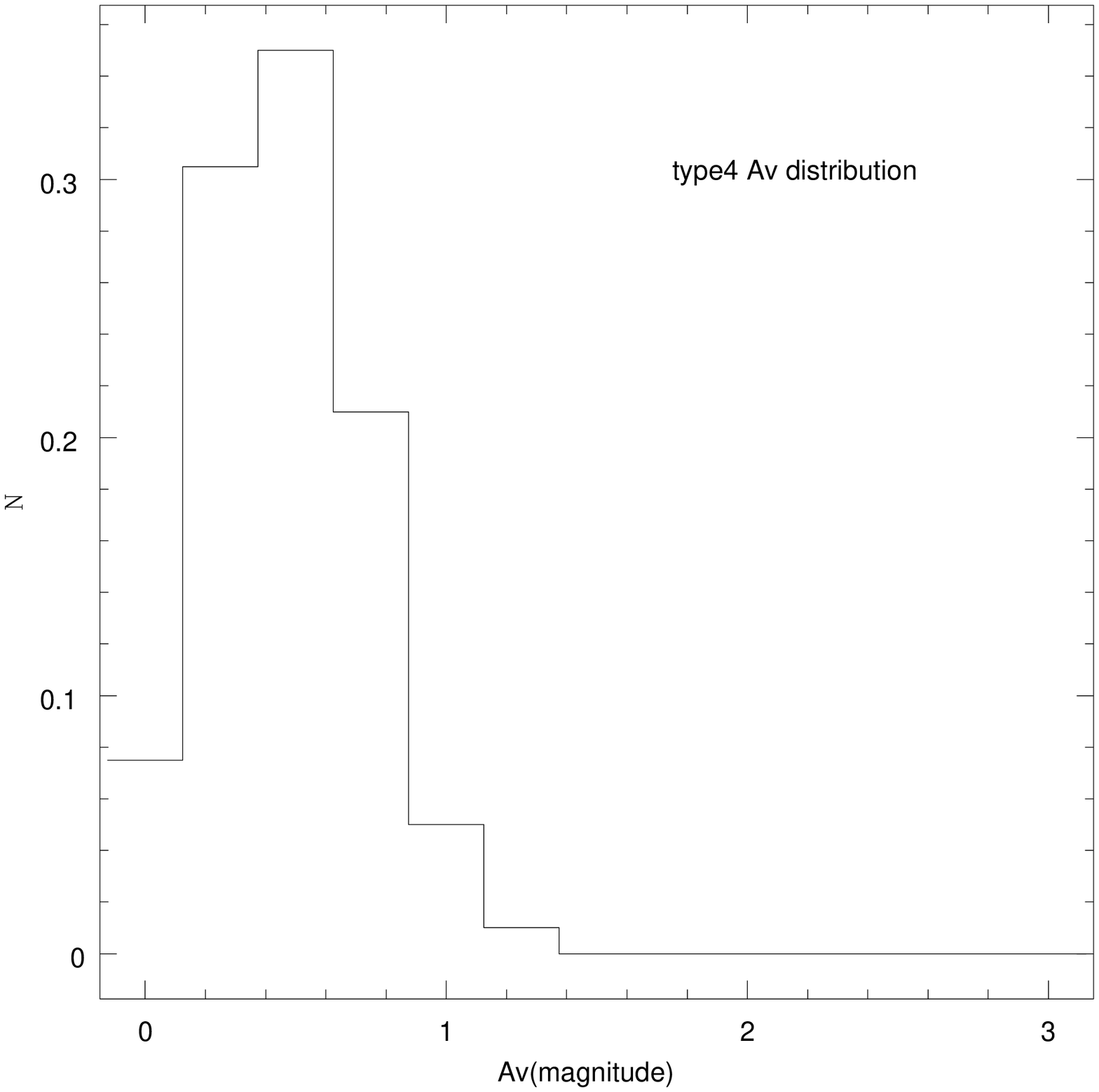}}
\caption{ Histogram showing extinction Av around the reddened type 4 stars
}
\end{figure}

\subsection{Type 3 stars}
This class of stars is one of the poorly populated class among the four classes identified. They are classified as showing periodic or quasi periodic photometric variations.
The fraction of stars identified are 6\% in the LMC (149 out of 2446) and 7.3\% in the SMC (78 out of 1056). Our cross identified
sample has 5.5\% in the LMC and 7.7\% in the SMC. This type of stars is found similarly populated in the L\&SMC as indicated by the above fractions.
The optical CMDs (Figures 1 \&2 ) show that these stars occupy similar location in the CMD and hence they are
likely to have similar evolutionary status in both the Clouds. 
The NIR CCDm for the type 3 stars in the LMC and the SMC are shown in Figure 10. 
In the NIR diagram, the type 3 stars are not located  as a homogeneous group.
Some stars can be found in the Be location and some stars are found along the MS.
They are more scattered in the LMC, when compared to the SMC. Some stars in the LMC
also show large NIR excess.  The histograms in figure 5 show that the type 3 stars occupy a larger
range of $(J-H)_0$ in the LMC. This is due to the presence of a few stars located close to the the MS in the NIR CCDm. The distribution in the $(H-K)_0$ suggests that the LMC stars have a larger range of $(H-K)_0$ values when compared to the SMC sample. The contribution comes from the blue stars located in the Be location as well as the stars with large NIR excess. 
The fact that they are not found in a coherent location,
may support the conclusion drawn by \citet {mennickent2002,mennickent2006} that these stars should not be linked to Be phenomenon at all. \citet{sabogal2005} found that type 3 stars have bimodal period distribution and that Double Periodic Variables (DPVs) are a small subgroup of type 3 stars. \citet{mennickent2006} studied some long period variables in the SMC and some of these are type 3 stars. They suggested that the type 3 stars are a mix of different kinds of stars. Their sample did not have any Herbig Ae/Be type stars, but has a number of binaries. \citet{mennickent2010} studied one of the type 3 stars from the above
sample using high resolution spectra and suggested that this star (smc3-1) could be a prototype of a small group of Magellanic Cloud wind-interacting A + B binaries. \citet{mennickent2010} studied the photometric variability of smc3-1 and found  a primary period of 238.1 days along with a complicated waveform suggesting ellipsoidal variability influenced by an eccentric orbit. This star also shows a secondary variability with an unstable periodicity that has a
mean value of 15.3 days. \citet {mennickent2010} suggested that this could be associated with non radial pulsations. 

We have spectra of type 3 candidates in both the LMC and the SMC. Spectra of 20 type 3 stars 
in the SMC were obtained, the details of these stars are tabulated in Table 4. We have also tabulated the
spectral type estimation of \citet {mennickent2006}, \citet{mennickent2010} (for smc3-1) as their spectra are of better resolution. We have tabulated the OGLEII number, optical magnitudes and NIR colours, reddening and radial velocity estimates by us and also by \citet{ mennickent2006} in Table 4.

We analyzed 15 spectra (blue) of type 3 stars in the SMC. The spectral line details of these stars are tabulated in Table 10.
Balmer lines H$\gamma$, H$\delta$, H$\xi$ are present as absorption lines. 
In a few cases, H$\beta$ is found in absorption. In 8 cases, H$\beta$ is seen in emission.
Lines of Hydrogen (H13, H12, H10, H9 and H8) are also present in absorption. He{\sc i} lines (4026, 4471\AA{}) 
are present in a few cases. We have 12 spectra in the red region. H$\alpha$ line is present in emission with varying $EW$ in all 
the spectra except in 3 cases where we can see emission in absorption line. The H$\alpha$ $EW$ ranges between
$-$82\AA{}  to +10\AA{}, where the positive $EW$ might suggest a filled in absorption.
Among these stars, smc3-1 has the highest H$\alpha$ $EW$ of $-$82 \AA, followed by smc3-20 and smc3-8,
with H$\alpha$ $EW$ of $-$37\AA{}  and $-$33\AA{} respectively. These stars are also found to have large values
of $(H-K)_0$ as seen from Table 4. The star smc3-1 is found to be an interacting binary with a circumbinary disc. It can be seen that H$\alpha$ $EW$ somewhat 
correlates with the $(H-K)_0$ magnitude. If we assume the large $(H-K)_0$ values indicate
NIR excess, then the above correlation might suggest the presence of circumstellar dust for these stars.
The reddening estimated towards these stars has a range between $E(B-V)$ = 0.04 - 0.61 mag. The reddening
values do not correlate with either the $(H-K)_0$ values or the H$\alpha$ $EW$. The large reddening values
observed in some stars might suggest association or proximity to star forming regions. The radial velocity estimates are comparable to those estimated by \citet{mennickent2006}, suggesting that
we do not detect any significant radial velocity variation. The star smc3-12 shows low radial velocity in both the measurements and hence is likely to be a member of the Galactic halo.

We analysed 12 spectra in the blue region and 9 spectra in the red region for LMC type 3 stars.
Balmer lines H$\gamma$, H$\delta$ and H$\xi$ are present as absorption lines. In 7 cases H$\beta$ 
is seen in absorption. In one case  H$\beta$ is seen in emission. 
Lines of Hydrogen (H13, H12, H10, H9 and H8) are also present in absorption. He{\sc i} lines (4026, 4471\AA{}) are
present and one star shows He{\sc i} 4026\AA{} in emission.  The spectral types, along with other details are tabulated in Table 5.

The spectral line details of these stars are tabulated in Table 11. H$\alpha$ line is present in emission with varying $EW$ in 5 stars. Three stars show H$\alpha$ in absorption, it may also be partially filled.
One star has relatively high $EW$ of $\sim$ $-$30\AA, whereas the rest have $EW$ less than $-$10\AA.
We find that the type 3 stars in the LMC have relatively less H$\alpha$ $EW$, when compared to those
in the SMC, keeping in mind that we only have a limited number of stars. We have spectrum of 
one star with large NIR excess, lmc3-20. This does not show any other emission other than H$_\alpha$ line. These stars do not show any correlation between $(H-K)_0$ values and H$\alpha$ $EW$. The reddening
range estimated for these stars is similar to those in the SMC, that is $E(B-V)$ = 0.05 - 0.60 mag. Thus, some stars might be located near star forming regions. 

Spectra of five type 3 stars in the SMC and one type 3 star in the LMC obtained from LCO100 are also analysed. The spectral line details of these stars are tabulated in Table 12. In the SMC,  H$\beta$ is found to be in absorption for 4 stars and in emission for one star. H$\gamma$ line is also seen in most cases as absorption. He{\sc i} lines 4471\AA{} and 4922\AA{} are seen in absorption in 4 spectra.
 
Using the photometric and the spectroscopic analysis of type 3 stars, we propose the following. 
These stars do not seem to have specific location in the NIR diagrams, supporting that 
these stars are unlikely to belong to one stellar population, or CBe stars. The fact that most 
of the stars show emission lines in the 
spectra (either H$\beta$ or H$\alpha$) suggest that they have circumstellar material. Most of the stars with the spectra are found to belong to A and F types, suggesting that the majority among type 3 are unlikely to be CBe stars. The fact that they correlate with the $(H-K)_0$ magnitude for some of the SMC candidates suggest the presence of dust in the disc. Some of the stars in the SMC are long periodic variables and one is found to be an
interacting binary. 3 stars in the LMC are found to belong to DPVs studied by \citet{ mennickent2005}. Thus, type 3 stars are likely to be a heterogeneous mix of stars belonging to various types of stars. 
  
\begin{figure}
\epsfxsize=10truecm
\centerline{\epsffile{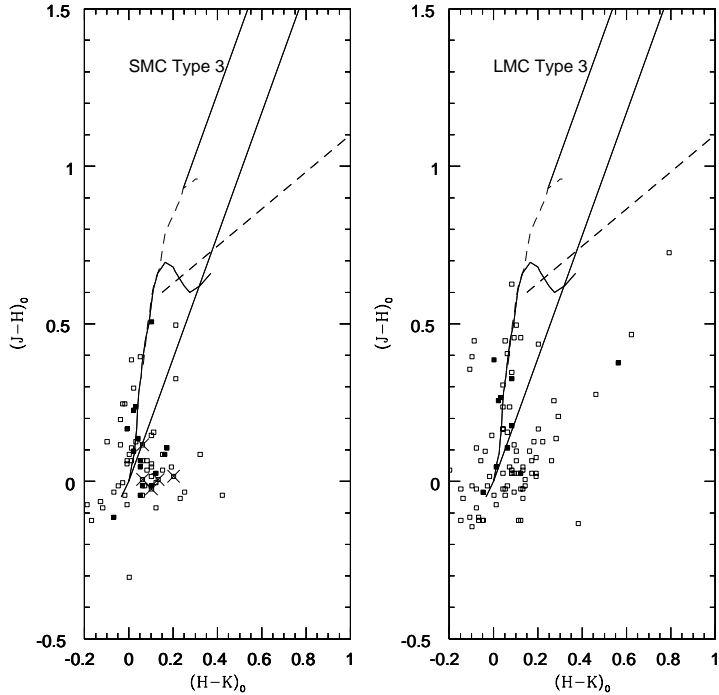}}
\caption{
NIR colour-colour diagram of the type 3 stars in the SMC and the LMC. Shaded points represent the stars with spectra from 1.5m CTIO 
and points shown with $\times$ represent stars with spectra from 2.5m LCO.
}
\end{figure}
\renewcommand{\thetable}{4}
\begin{landscape}
\begin{table*}
\begin{center}

\caption{Type 3 stars in the SMC present in our sample along with optical and infrared magnitude and colours. Spectral type determined in this study is given in column 7 and  spectral type determined by \citet{mennickent2010}  for the star smc3-1 and \citet{mennickent2006} for the rest of the stars is given in column 8. Our determination of radial velocity along with the radial velocity determined by \citet{mennickent2006} are also given. N is the number of lines averaged for radial velocity determination.}
\end{center}
\begin{tabular}{lrrrrrrrrrrrr}
\hline
\hline
Star & OGLE-name & $(B-V)_0$ & $V_0$ & $(H-K)_0$ & $(J-H)_0$ &Sp.type& Sp.class & $E(B-V)$&v$\pm\sigma_{v}$&N&v$\pm\sigma_{v}$&N\\
&&&&&&(This study)&Menn.(2006)&&\multicolumn{2}{c}{(This study)}&\multicolumn{2}{c}{Menn.(2006)}\\
\hline
&Spectra from 1.5m CTIO&&&&&&\\
\hline
smc3-1&OGLE004336.91-732637.7&-0.19&13.04&0.13&0.02&A3-A5&A+B&0.37&117$\pm32$&6&114$\pm31$&15\\
smc3-2&OGLE004454.66-730802.9&-0.16&13.44&-0.01&0.15&B8-A1&B7-9IIIe&0.44&130$\pm36$&7&137$\pm26$&7\\
smc3-3&OGLE004554.14-731404.3&0.39&15.09&0.04&0.25&A7-F0&F0 Ib-II&0.13&164$\pm37$&5&174$\pm23$&7\\	
smc3-5&OGLE004750.14-731316.4&0.07&14.85&0.05&0.06&B8&B7-8 III&0.20&167$\pm69$&6&139$\pm5$&7\\
smc3-6&OGLE004921.41-725844.9&0.20&14.23&0.06&0.16&A7-F0&A5 II&0.12&199$\pm30$&8&191$\pm23$&13\\
smc3-7&OGLE005025.64-725807.1&0.17&13.63&0.01&0.19&A7-F0&F4IV&0.12&150$\pm29$&7&127$\pm15$&10\\
smc3-8&OGLE005043.44-732705.3&-0.24&13.51&0.13&0.04&--&WR&0.05&192$\pm58$&3&179$\pm26$&4\\
smc3-9&OGLE005100.18-725303.9&-0.15&13.23&0.12&0.53&B5-B8&B1II-IIIe&0.12&170$\pm77$&4&172$\pm32$&12\\
smc3-12&OGLE005359.22-723508.9&-0.25&13.27&-0.01&0.04&&A3III&0.36&66$\pm23$&7&54$\pm17$&12\\
smc3-15&OGLE005745.25-723532.1&-0.27&13.43&-0.05&-0.09&B2-B5&B1II-III&0.13&220$\pm23$&7&195$\pm37$&22\\
smc3-18&OGLE010000.78-725522.9&-0.32&13.10&0.02&-0.11&&B8-9IIIe&0.37&240$\pm24$&6&222$\pm32$&14\\
smc3-20&OGLE010451.21-724646.9&-0.09&13.29&0.15&0.08&B5-B8&B1Ib-IIe&0.04&172$\pm9$&8&179$\pm32$&16\\
smc3-21&OGLE010452.99-715918.8&-0.21&14.09&0.09&0.04&B2-B5&B2IIIe&0.04&132$\pm33$&5&143$\pm24$&15\\
smc3-14&OGLE005520.27-723710.1&0.77&12.75&&&A0-A3&A0Ibe&0.61&122$\pm37$&5&141$\pm16$&2\\
smc3-16&OGLE005812.58-723048.5&-0.42&14.62&&&B8-A0&B0IIIe&0.20&166$\pm40$&4&164$\pm24$&4\\
\hline
&Spectra  from 2.5m LCO &&&&&&\\
\hline
smc3-30&OGLE005710.71-722550.2&0.17&15.48&0.09&0.17&&&0.04\\
smc3-31&OGLE005617.49-730005.1&-0.55&14.44&0.00&-0.12&&&0.53\\
smc3-32&OGLE004652.17-731409.2&-0.27&15.36&0.20&0.01&&&0.20\\
smc3-33&OGLE005059.66-725648.3&-0.13&15.22&0.15&0.03&&&0.12\\
smc3-35&OGLE005217.74-725627.9&-0.11&16.15&0.13&0.03&&&0.04\\
\hline
\end{tabular}
\end{table*}
\end{landscape}

\renewcommand{\thetable}{5}
\begin{table*}
\begin{center}
\caption{Type 3 stars in the LMC present in our sample along with our determinations for spectral type, magnitude, colour and radial velocity. N is the number of lines averaged for radial velocity determination.}
\end{center}
\begin{tabular}{lrrrrrrrrr}
\hline
\hline
Star & OGLE-name &  $(B-V)_0$ & $V_0$ & $(H-K)_0$ & $(J-H)_0$ & Sp.type & $E(B-V)$&v$\pm\sigma_{v}$&N\\
\hline
&Spectra obtained from 1.5m CTIO&&&&&&\\
\hline
lmc3-1&	OGLE05005236-685803.7&	0.61&13.84&0.05&0.31&F9&0.05&149$\pm2$&2\\
lmc3-5&	OGLE05030370-690615.0&	0.10&	13.75&0.03&	0.07&	A9-F0&0.13&193$\pm33$&8\\
lmc3-20&OGLE05141821-691235.0&	0.42&	14.43&0.59&	0.43&	F6&0.04&226$\pm20$&5\\
lmc3-23&OGLE05164754-694415.2&	0.47&14.74&0.10&	0.35&	F3&	0.12&264$\pm33$&7\\
lmc3-24&OGLE05174442-692033.3&	-0.09&12.75&-0.03&-0.01&	B8-A1&0.12&301$\pm13$&10\\
lmc3-30&OGLE05194782-693912.3&	0.21&	13.05&0.05&	0.07&	F6&0.29&166$\pm43$&4\\
lmc3-33&OGLE05203226-694224.2&	0.14&	13.61&0.04&	0.09&	F6&0.44&262$\pm20$&6\\
lmc3-37&OGLE05225847-692621.0&	0.17&	13.91&0.02&	0.23&	F6&0.28&250$\pm99$&4\\
lmc3-39&OGLE05240201-694920.5&	-0.78&11.45&-0.03&	0.24&	B2-B3&0.60&245$\pm22$&8\\
lmc3-44&OGLE05265249-693317.2&-0.10&	14.62&0.15&	0.08&	--&0.05&-&-\\
lmc3-12&OGLE05084863-684315.6&-0.22&14.31&&&A6&0.21&192$\pm21$&9\\
lmc3-49&OGLE05293898-693448.0&-0.78&14.30&&&B2-B3&0.60&266$\pm22$&9\\
\hline
\end{tabular}
\end{table*}

\subsection{Type 2 stars}
The NIR CCDm for the type 2 stars in the LMC and the SMC are shown in Figure 11.
Relatively more type 2 stars exists in the SMC (14.5\%) when compared to the LMC (6\%). The cross identified sample also shows the overpopulation in the SMC (15.6\%) when compared to the LMC (5.6\%).
It is clear that the SMC has a larger fraction of this type when compared to the LMC. Thus, the identification of the nature of these stars might throw some insight into the reason for the above difference. This type is also one of the poorly populated sample among the 4 types of Be candidates. Majority of these stars show NIR excess 
similar to the Galactic Be stars. There is no significant difference in the NIR properties
between the LMC and the SMC type 2 stars. There are a few stars which are located along the MS. This scatter
is relatively more in the LMC, than in the SMC. The histograms presented in figure 5 show that the type 2 stars in the SMC are slightly redder in $(J-H)_0$ than those in the LMC, whereas the colour distribution in $(H-K)_0$ is very similar in both the galaxies.  None of the type 2 stars show significant NIR excess, which is typical for pre-MS stars. \citet{mennickent2002} suggested that type 2 stars
could be Be+WD binary stars or accreting pre-MS stars, based on the light curve analysis of these stars.
The NIR properties suggest that the most of the LMC type 2 stars and  probably all the SMC type 2 stars are similar to the Galactic Be stars. They do not show the NIR properties typical of the pre-MS stars. 
 
We obtained spectra of 20 type 2 stars in the SMC (12 in blue region and 8 in the red region) and the spectral classification suggests that most of them are early type stars. 
Stars observed are tabulated in Table 6 along with
their identification number, optical magnitudes, NIR colours and the estimated spectral types.
The identified spectral lines and the equivalent width ($EW$) of the prominent lines are tabulated in Table 10.

Balmer lines H$\gamma$, H$\delta$  and lines of Hydrogen (H13, H12, H10, H9, H8)
are present in absorption. He{\sc i} lines (4009, 4026, 4471, 4390\AA{}) are present in a few cases and 5 out of 12 stars show H$\beta$ emission. 4 spectra show no emission or absorption corresponding to H$\beta$, probably due to filled up absorption. In one case there is absorption corresponding to H$\beta$. Among the 8 red spectra, H$\alpha$ line is present in emission with varying $EW$ (-34\AA{} to +1.3\AA{}) in  7 spectra. \citet{martayan2007} also found a range of H$\alpha$ $EW$ among the CBe stars in the LMC. Of the 12 stars, therefore, we find that emission lines are present in at least 7 stars. Thus, most of the stars have circumstellar material and  these are early type stars. Spectra of 9 stars in the LMC and 4 stars in the SMC obtained from LCO100 are also analysed. The spectral types, along with other details are tabulated in Table 6 and 7. The spectral line details of these stars are tabulated in Table 12. In the LMC, 5 stars show H$\beta$ in emission. He{\sc i} 4922\AA{} is found in absorption in 7 stars. In the SMC H$\beta$ is found in absorption in 2 stars and emission in 2 stars. 

The reddening values for these stars show that about half of them have very low reddening of $E(B-V)$ = 0.04 and the half have values ranging up to 0.45 mag. They are unlikely to be pre-MS stars due to lack of large NIR excess. Among the 12 stars classified, 10 stars belong to early B-type. The difference in the variability might give
clues to their nature, as these stars show sudden luminosity jumps. These type of stars are found to be more in the SMC than in the LMC. Their location in the NIR CCDm suggests that these stars in the SMC appear like a homogeneous group, whereas the LMC stars do not appear as a homogeneous group. The location of these stars in the optical CMDs are similar for the LMC and the SMC. If these are early type stars, then it might appear that the SMC sample might have a larger fraction of the more massive Be stars when compared to the LMC as reported by \citet{martayan2010} and \citet{bonanos2010} for early Be and Oe stars. This could happen due to the lower metallicity of the SMC. These stars could be very young Be stars, or could be Be stars in binaries as suggested by \citet{mennickent2002}. Therefore, it will be interesting to study these type 2 stars in detail in the LMC also, in order to compare them with the SMC.

\begin{figure}
\epsfxsize=10truecm
\centerline{\epsffile{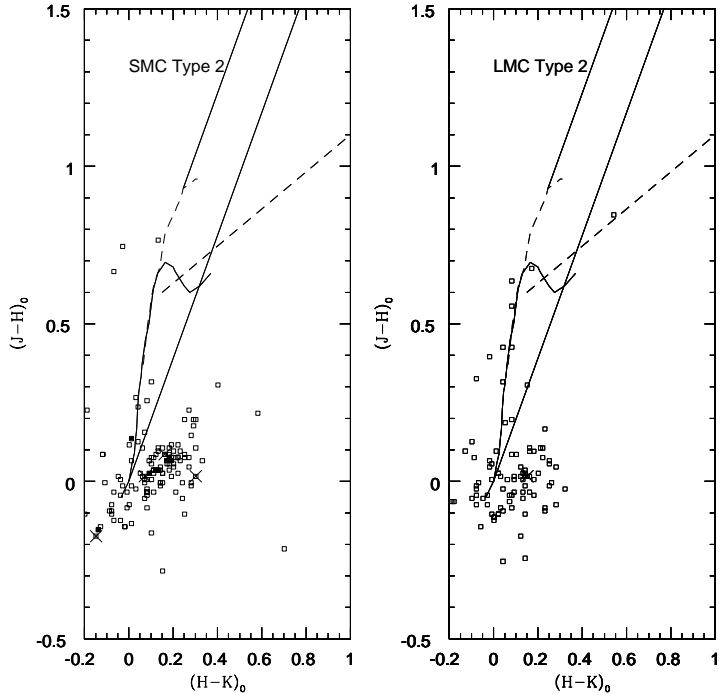}}
\caption{ NIR colour-colour diagram of the type 2 Be candidates in the  LMC and the SMC. Shaded points represent the stars with spectra from 1.5m CTIO and points shown with $\times$ represent stars with spectra taken from 2.5m LCO.
}
\end{figure}
\renewcommand{\thetable}{6}
\begin{table*}
\begin{center}
\caption{Type 2 stars in the SMC present in our sample along with our determinations for spectral type, magnitude, color and radial velocity. N is the number of lines averaged for radial velocity determination.}
\end{center}
\begin{tabular}{lrrrrrrrrr}
\hline
\hline
Star & OGLE-name &  $(B-V)_0$ & $V_0$ & $(H-K)_0$ & $(J-H)_0$ & Sp.type & $E(B-V)$&v$\pm\sigma_{v}$&N\\
\hline
&Spectra obtained from 1.5m CTIO&&&&&&\\
\hline
smc2-4&OGLE004721.86-730650.1&-0.31&13.15&0.16&0.04&B0-B2&0.28&188$\pm11$&4\\
smc2-6&OGLE004848.83-730620.1 & -0.08&14.25&0.22&0.12&B0-B2&0.04&160$\pm15$&2\\
smc2-8&OGLE004849.49-724800.0 & -0.15&14.79&0.15&0.09&B0-B2&0.04&187$\pm26$&5\\
smc2-12&OGLE005224.40-724038.6&-0.04&13.90&0.12&0.08&B0-B2&0.04&115$\pm24$&6\\
smc2-14&OGLE005251.97-723508.5&-0.20&14.50&0.16&0.09&B0-B3&0.04&218$\pm32$&5\\
smc2-19&OGLE005614.45-724053.2&0.17&13.96&0.03&0.16&A5-A7&0.13&131$\pm16$&8\\
smc2-23&OGLE005806.05-723544.5&-0.21&14.21&0.18&0.07&B0-B3&0.20&192$\pm18$&5\\
smc2-27&OGLE010025.10-724632.2&-0.15&14.57&-0.11&-0.10&B0-B2&0.04&205$\pm19$&5\\
smc2-9&OGLE004938.01-730610.0&-0.17&14.30&&&B0-B3&0.04&149$\pm20$&3\\
smc2-20&OGLE005618.51-722645.2&-0.40&13.77&&&A0-A3&0.45&108$\pm35$&7\\
smc2-29&OGLE010401.11-723311.1&-0.06&12.98&&&B0-B3&0.04&167$\pm25$&8\\
smc2-32&OGLE010447.25-722559.4&-0.08&14.65&&&B0-B2&0.04&85$\pm37$&6\\
\hline
&Spectra obtained from 2.5m LCO &&&&&&\\
\hline
smc2-38&OGLE004236.69-733033.1&-0.27&15.60&0.09&0.04&&0.12\\
smc2-39&OGLE004304.50-730206.3&-0.34&14.92&0.13&0.02&&0.37\\
smc2-40&OGLE004327.71-731653.9&-0.42&15.26&0.27&-0.05&&0.37\\
smc2-45&OGLE004521.57-731717.7&-0.55&14.53&-0.19&-0.26&&0.44\\
\hline
\end{tabular}
\end{table*}
\renewcommand{\thetable}{7}
\begin{table*}
\begin{center}
\caption{Type 2 stars in the LMC present in our sample along with our determinations of magnitude and colours.}
\end{center}
\begin{tabular}{lrrrrrr}
\hline
\hline
Star & OGLE-name &  $(B-V)_0$ & $V_0$ & $(H-K)_0$ & $(J-H)_0$ &  $E(B-V)$\\
\hline
&Spectra obtained from 2.5m LCO &&&&&\\
\hline
lmc2-1&OGLE050017.58-692749.9&-0.11&14.71&--&0.59&0.13\\
lmc2-2&OGLE050039.16-692003.9&-0.16&15.02&-0.02&--&0.13\\
lmc2-3&OGLE050053.02-692011.5&-0.15&13.84&-0.02&--&0.13\\
lmc2-4&OGLE050103.77-691746.7&-0.15&14.50&0.14&0.65&0.13\\
lmc2-5&OGLE050116.74-692027.7&-0.25&14.66&0.08&0.46&0.13\\
lmc2-6&OGLE050306.60-691807.5&-0.13&14.89&-0.02&--&0.13\\
lmc2-7&OGLE050340.10-691530.9&-0.13&13.96&--&0.30&0.13\\
lmc2-8&OGLE050350.56-690223.0&-0.19&14.63&0.15&0.57&0.13\\
lmc2-9&OGLE050412.28-685021.7&-0.18&14.99&-0.02&--&0.13\\
\hline
\end{tabular}
\end{table*}
\subsection{Type 1 stars}
The NIR CCDm for the type 1 stars in the LMC and the SMC are shown in Figure 12.
Type 1 stars are relatively more numerous in the LMC (24\%), when compared to the SMC (13\%). The cross-correlated sample has 24\% in the LMC and 10\% in the SMC. We have cross identified a smaller sample of this
type in the SMC. Type 1 stars are the second largest population among the four types and forms one fourth of the identified sample in the LMC. These stars are populated more in the LMC, when compared to the SMC, and the ratio is almost twice. Thus, it will be interesting to find the reason for their overabundance in the LMC or their under abundance in the SMC. Majority of type 1 stars occupy similar location as Galactic Be stars in the NIR CCDm. These stars do not have large NIR excess.  The histograms shown in figure 5 suggest that the distribution in $(J-H)_0$ as well as $(H-K)_0$ are very similar in L\&SMC, except for the large number present in the LMC. Thus, these stars are found to have similar NIR properties in the LMC and the SMC. \citet{mennickent2002} had stated that the type 1 could be similar to Galactic Be stars. 

In the SMC type 1 stars, we have analyzed 17 spectra in the blue region and 13 spectra in the red region. The spectral line details of these stars are tabulated in Table 10.
Balmer lines H$\beta$, H$\gamma$, H$\delta$ and
lines of Hydrogen (H13, H12, H10, H9, H8) are present in absorption. He{\sc i} lines (4009, 4026, 4471, 4390\AA{}) are present with varying intensities in absorption.
7 (stars 7, 18, 26, 28, 31, 37 and 38) out of 17 stars show H$\beta$ emission. Filled in profile is seen in H$\delta$ 
in a few cases. H$\alpha$ line is present in emission with varying EW in all the spectra. The estimated range is between $-$33.8 and $-$0.94\AA{}. The spectral types, along with other details are tabulated in Table 8.  These stars are found to belong to early B type stars, which is consistent with their location in the brighter part of the optical CMD (Figure 2). lmc1-55 is the only type 1 star in the LMC for which we could estimate the spectral type, and is found to belong to B0-B2.

Spectra of 17 stars in the LMC and 12 stars in the SMC obtained from LCO100 are also analysed (Tables 8 and 9). In the LMC, Balmer lines H$\gamma$ and  H$\beta$ are found in absorption for 14 stars in our sample while in 3 stars they are seen as  emission lines. As tabulated in Table 12 quite number of He{\sc i} lines (4387, 4471, 4713, 4922\AA{}) are present in  absorption in many cases. 8 stars are found to have 5169\AA{} Fe{\sc ii} line in absorption. In the SMC out of 12 spectra, 6 show H$\beta$ in absorption and 5 in emission. He{\sc i} lines are also present in majority of the cases in absorption. Fe{\sc ii} 5169\AA{} is found in emission in four stars. Our analysis shows that these stars are B-type stars though we are unable to identify the exact spectral sub class. Many of these stars are found to lie  in the CBe-location in the  CCDm and a few are found on the MS above CBe location. Since H$\beta$ is found in emission for 5 stars, it can be assumed that H$\alpha$ also will be in emission. Hence we conclude that half the stars show H$\alpha$ in emission.

Most of the type 1 stars in the SMC are found to show emission features, whereas the LMC stars do not show clear emission features. We could not estimate the spectral type of these stars in the LMC because of poor signal to noise ratio. Their location in the optical CMD (Figure 1) suggests that the majority could belong to later spectral types. Hence it is probable that type 1 stars in the LMC and the SMC may belong to different spectral types, but with properties similar to CBe stars.
   
\begin{figure}
\epsfxsize=9truecm
\centerline{\epsffile{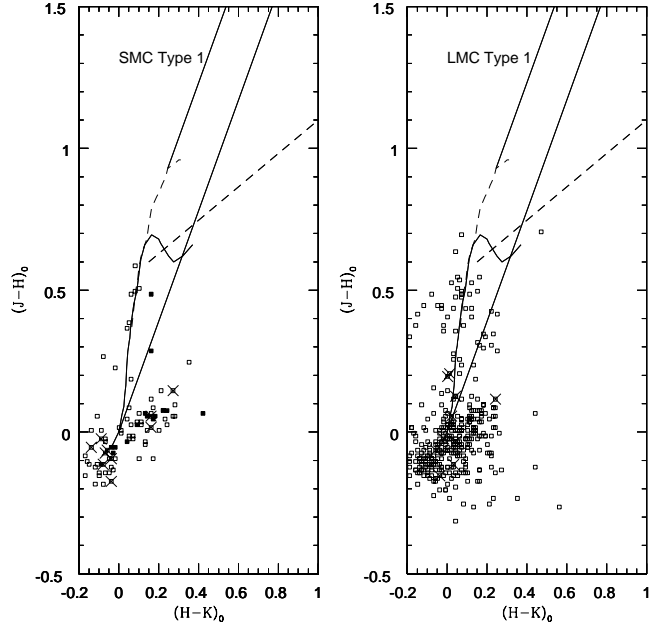}}
\caption{ NIR colour-colour diagram of the type 1 stars in the  LMC and the SMC. Shaded points represent the stars with spectra from 1.5m CTIO and points shown with $\times$ represent stars with spectra from 2.5m LCO.
}
\end{figure}
\renewcommand{\thetable}{8}
\begin{table*}
\begin{center}
\caption{Type 1 stars in the SMC present in our sample along with our determinations for spectral type, magnitude, color and radial velocity. N is the number of lines averaged for radial velocity determination.}
\end{center}
\begin{tabular}{lrrrrrrrrr}
\hline
\hline
Star & OGLE-name  & $(B-V)_0$ & $V_0$ & $(H-K)_0$ & $(J-H)_0$ & Sp.type &  $E(B-V)$&v$\pm\sigma_{v}$&N\\
\hline
&Spectra obtained from 1.5m CTIO&&&&&&\\
\hline
smc1-7 & OGLE005059.21-724357.3 &-0.06 &14.67 & 0.19 &0.34&B0-B2&0.04&205$\pm2$&2\\	
smc1-10 & OGLE005224.40-724038.6 &-0.04 &13.90 & 0.12 &0.09&B0-B3&0.04&149$\pm30$&6\\
smc1-11 & OGLE005227.51-732001.2 &-0.28 &12.30 & -0.03 &-0.09&B0-B2&0.28&200$\pm13$&8\\
smc1-17 & OGLE005504.55-724637.3 &-0.31 &13.53 & -0.09&-0.12&B0-B3&0.20&163$\pm24$&5\\
smc1-18 & OGLE005614.63-723755.1 &-0.08 &14.18 & 0.19&0.07&B0-B3&0.13&247$\pm38$&5\\
smc1-19 & OGLE005641.86-724425.4 &-0.24 &14.33 & 0.44&0.09&B0-B3&0.13&167$\pm28$&6\\
smc1-21 & OGLE005916.06-722100.3 &-0.24 &13.88 & 0.22&0.07&B0-B3&0.20&223$\pm51$&3\\
smc1-26 & OGLE010058.69-723049.9 &-0.13 &13.82 & 0.17&0.11&B0-B3&0.04&158$\pm23$&7\\
smc1-28 & OGLE010213.80-722213.0 &-0.01 &14.24 & 0.21&0.11&B0-B3&0.04&176$\pm18$&3\\
smc1-31 & OGLE010542.57-722747.3 &0.09 &14.91 & 0.18 &0.15&B0-B3&-0.04&170$\pm28$&5\\
smc1-36 & OGLE010807.38-721932.6 &-0.23 &14.04& 0.19&0.54&B0-B2&0.05&179$\pm24$&9\\
smc1-37 & OGLE010809.52-721556.8 &-0.23 &14.34& 0.07&0.02&B0-B3&0.05&167$\pm32$&7\\
smc1-38 & OGLE010825.82-722327.2 &-0.08 &13.82 & 0.27&0.13&B0-B3&0.05&228$\pm10$&2\\
smc1-2 & OGLE003918.20-733656.6  & 0.20 & 13.66 & &&F0&0.29&200$\pm16$&5\\
smc1-23&OGLE010041.93-723028.6&-0.63&14.59&&&B0-B2&0.45&238$\pm35$&7\\
smc1-24 & OGLE010043.94-722604.8 &-0.64 &14.37 & &&B0-B3&0.45&188$\pm14$&5\\
smc1-39&OGLE005235.60-723751.7&--&--&&&B0-B2&--&188$\pm19$&5\\
\hline
&Spectra obtained from 2.5m LCO &&&&&&\\
\hline
smc1-40&OGLE003631.42-733917.8&-0.36&15.76&-0.02&-0.05&&0.13\\
smc1-44&OGLE004225.27-731718.2&-0.52&14.83&-0.07&-0.16&&0.37\\
smc1-47&OGLE004626.57-731929.3&-0.13&15.22&-0.09&-0.03&&0.20\\
smc1-50&OGLE004701.71-731041.9&-0.39&14.64&-0.09&-0.14&&0.28\\
smc1-52&OGLE004801.80-731057.0&-0.32&14.84&-0.08&-0.10&&0.28\\
smc1-53&OGLE004803.29-730722.2&-0.27&16.22&0.19&0.07&&0.04\\
smc1-56&OGLE004916.20-724941.2&-0.05&15.34&-0.12&-0.03&&0.12\\
smc1-57&OGLE004930.81-731236.5&-0.16&15.39&-0.04&-0.04&&0.12\\
smc1-59&OGLE004958.39-725750.8&-0.08&15.49&0.18&0.08&&0.12\\
smc1-60&OGLE005018.69-725524.3&0.00&15.62&0.29&0.17&&0.12\\
\hline
\end{tabular}
\end{table*}
\renewcommand{\thetable}{9}
\begin{table*}
\begin{center}
\caption{Type 1 stars in the LMC present in our sample along with our determinations for  magnitude, color and radial velocity. N is the number of lines averaged for radial velocity determination.}
\end{center}
\begin{tabular}{lrrrrrrrr}
\hline
\hline
Star & OGLE-name  & $(B-V)_0$ & $V_0$ & $(H-K)_0$ & $(J-H)_0$ &    $E(B-V)$&v$\pm\sigma_{v}$&N\\
\hline
&Spectra obtained from 1.5m CTIO&&&&&&\\
\hline
lmc1-55&OGLE051747.85-690908.6&-0.23&13.31&0.11&0.13&0.36&287$\pm38$&7\\
\hline
&Spectra obtained from 2.5m LCO &&&&&\\
\hline
lmc1-1&OGLE050122.35-690004.9&-0.10&15.07&--&0.80&0.05\\
lmc1-2&OGLE050131.42-684251.0&0.32&14.09&0.07&0.18&0.05&107$\pm22$&3\\
lmc1-3&OGLE050132.69-692845.8&-0.25&14.67&0.05&-0.09&0.13&263$\pm31$&2\\
lmc1-4&OGLE050147.90-692424.1&0.43&13.99&0.02&0.22&0.13\\
lmc1-5&OGLE050151.04-692109.6&0.11&13.94&0.04&0.08&0.13&117$\pm38$&2\\
lmc1-6&OGLE050158.32-684315.5&0.29&15.34&0.09&0.52&0.05\\
lmc1-7&OGLE050221.24-690451.6&-0.21&14.38&-0.04&-0.04&0.13&203$\pm32$&2\\
lmc1-9&OGLE050127.92-685947.7&0.42&14.31&0.03&0.23&0.13\\
lmc1-10&OGLE050234.22-683939.8&-0.05&14.14&-0.02&--&0.13\\
lmc1-11&OGLE050320.42-685315.7&-0.35&14.25&0.18&0.19&0.28&254$\pm28$&2\\
lmc1-12&OGLE050438.96-691624.6&-0.26&13.61&0.16&0.36&0.13&242$\pm2$&2\\
lmc1-13&OGLE050426.31-690714.4&-0.32&14.29&--&-0.21&0.36\\
lmc1-15&OGLE050447.46-690552.3&-0.15&14.24&-0.07&--&0.36\\
lmc1-114&OGLE050151.61-690152.8&-0.36&15.68&0.21&0.05&0.37\\
lmc1-115&OGLE050428.01-690927.5&-0.03&15.85&-0.07&--&0.36\\
lmc1-116&OGLE050521.34-691550.1&-0.35&14.48&-0.07&--&0.36\\
lmc1-117&OGLE053319.88-701213.1&-0.01&16.39&0.07&0.02&0.05\\
\hline
\end{tabular}
\end{table*}

\renewcommand{\thetable}{10}
\begin{table*}
\begin{center}
\caption{List of spectral lines observed in the sample of Be candidate stars in the SMC (type 1, 2 and 3 as defined by \citet{mennickent2006}) considered in the present study along with their equivalent widths in \AA.  ``\# ''indicates that there is no spectra available for these stars in the red region. ``\#\#'' indicates that there is no spectra available for these stars in blue region. ``ea'' indicates emission in absorption line in H$\alpha$. Numbers mentioned in the brackets indicate the number of spectra in the red region for that particular star.}
\end{center}
\begin{tabular}{lrrrrrrrrrl}
\hline
\hline
Star&	CaII&	H$\varepsilon$& 	He{\sc i}&	H$\delta$ & 	H$\gamma$ &	He{\sc i}&	He{\sc i}&	H$\beta$&H$\alpha$&Comm\\
\hline
smc1-2&	8.27&6.97&--&2.34&2.84&	--& --& --& --&\\			
smc1-7& --&  --&1.16&	1.01&	--&--&	0.34&	-4.57&	-32.83&\\
smc1-10&0.67&	1.64&	0.62&	1.67&	2.27&	0.88&	0.69&--&-8.23&\\
smc1-11	&0.70	&2.71	&0.63 &	1.65 &	1.83 &	0.57 &	0.72&	1.8&&\#\\
smc1-16&&&&&&&&&-8.95&\#\#\\	
smc1-17&0.30&	2.28&	0.65&	1.27&	1.85&	0.43&	0.29&--&&\#\\		
smc1-18	&--&0.90&0.7&	1.17&	--&--&ea&-2.91&-30.66&\\
smc1-19	&--&1.92&0.58&1.22	&--&--&0.4&--&&\#\\				
smc1-21	&0.50&2.77&0.87&1.91&1.89&0.46&0.73&1.85&-31.28&\\
smc1-23&0.89&2.89&0.9&2.49&2.36&0.27&1.13&2.153&&\#\\	
smc1-24	&0.48	&1.9&	0.63&1.21&1.27	&--&0.28&--&&\#\\
smc1-25&&&&&&&&&-24.28&\#\# (4)\\			
smc1-26	&--&1.50&0.82&0.90&	--&0.34&0.45&	-2.23&	-16.64&\\
smc1-28	&--&	2.09&	0.63&	0.842&	--&	0.55&	0.38&	-2.91&	-23.42&\\
smc1-31	&0.70	&1.85&	0.81	&1.24&--&--&--&-3.70&&\#\\	
smc1-36	&--&3.12&0.94&2.18	&2.58&	0.58	&0.82&	2.24	&-0.94&\\
smc1-37	&0.31	&2.25	&0.82	&1.95	&1.60&--&0.49	&-1.40&-12.56&\\
smc1-38	&--&--&	0.40	&--	&-1.03	&--&--&	-4.27&-31.74&\\
smc1-39 &0.48&3.92&1.10&2.20&2.36&0.51&0.40&0.19&-8.96& \\							
smc2-4&	0.44&	1.34&	0.50&	1.41&	-- &   --&0.34	&-3.53	&-33.84&\\
smc2-6	&--&	1.26	&--&--&--&--&--&-2.86&	-21.08&\\
smc2-8	&0.74&	3.15	&1.4&	2.07&1.9&--&0.87&--&ea&\\					
smc2-9&--&--&0.86&1.14&1.85&--&0.46&--&-13.87&\\
smc2-12&0.30&	2.50&0.4&1.76&	1.71&0.32&ea&ea&&\#\\				
smc2-14	&0.58&	1.45	&0.50&--&--&0.32&0.85&-2.59&-17.91&\\	
smc2-19	&3.98&	7.96&--&4.87&	5.97&--&--&4.79&&\#\\
smc2-20&1.32&7.05&0.09&6.10&6.88&--&0.08&6.32&&\#\\	
smc2-23	&--&2.06&0.72	&1.47	&1.93&0.33&085&0.93&&\#\\
smc2-27&0.88&4.25&1.32&3.01&3.31&0.62&1.09&2.74&1.31&\\\
smc2-29&0.99&1.99&0.66&1.36&-0.45&0.35&0.63&-2.90&-24.29&\\
smc2-32&0.12&1.49&1.0&1.23&-0.12&0.13&0.91&-3.08&-25.67&\\							
smc3-1	&2.38	&4.54	&--&--&--&--&--&-8.00	&-82.36&\\
smc3-2	&0.8&	6.25&	--&	3.28&3.4&0.7&--&--&-5.95&(2)\\
smc3-3	&5.65&6.63&--&4.55&3.16&0.55&--&6.09&1.64&(3)\\
smc3-4&&&&&&&&&9.90&\#\# (3)\\
smc3-5	&--&7.28&--&	5.30&	9.24	&--&ea&	6.01&&\#\\	
smc3-6	&5.19	&8.96&	--&	6.93&	5.44	&--&--&	5.72&&\#\\	
smc3-7	&5.52&	8.36&--&5.69&	5.54&--&--&6.37&&\#\\	
smc3-8	&0.77&	0.73&--&-2.09&-1.76&--&--&-7.22&-33.28&(3)\\
smc3-9	&--&	0.72&	0.57&	-0.34&--&0.34&0.5&-2.12&	-13.83&\\
smc3-10&&&&&&&&&4.68&\#\#\\
smc3-12	&2.74	&6.28&0.5&5.31	&4.54	&--&0.5&	4.43&&\#\\
smc3-14&0.81&2.49&--&2.23&1.99&0.30&0.07&-0.57&-13.67&\\	
smc3-15	&0.79	&2.64&	0.91&	1.55&2.21&0.46&	0.75&	1.52&&\#\\	
smc3-16	&0.59	&0.65	&0.63	&ea&--&--&0.4&-2.17&-10.37&\\
smc3-17&&&&&&&&&-6.20&\#\#\\
smc3-18	&1.20&	3.38&--&2.67	&1.76&--&--&-2.43&&\#\\	
smc3-20&0.65&1.85&0.25&	0.82	&-0.24&0.52&0.63&-3.27&-37.26&\\
smc3-21	&--&2.24&0.77	&1.72&	2.60&--&0.94&	-0.59&	-9.87&\\

\hline
\end{tabular}
\end{table*}

\renewcommand{\thetable}{11}
\begin{table*}
\begin{center}
\caption{List of spectral lines observed in the sample of Be candidate stars in the LMC (type 1, 2 and 3 as defined by \citet{mennickent2006}) considered in the present study along with their equivalent widths in \AA. ``\# ''indicates that there is no spectra available for these stars in the red region. ``\#\#'' indicates that there is no spectra available for these stars in blue region. ``ae'' indicates absorption in emission line in H$\alpha$.} 
\end{center}
\begin{tabular}{lrrrrrrrrrl}
\hline
\hline
Star&	CaII&	H$\varepsilon$& 	He{\sc i}&	H$\delta$ & 	H$\gamma$ &	He{\sc i}&	He{\sc i}&	H$\beta$&H$\alpha$&Comm\\
\hline

lmc1-8&&&&&&&&&3.60&\#\#\\
lmc1-12&&&&&&&&&-16.12&\#\#\\
lmc1-52&&&&&&&&&-11.33&\#\#\\
lmc1-55&0.44&2.98&0.65&0.38&0.34&0.13&0.72&-1.64&-13.27&\\
lmc1-60&&&&&&&&&2.43&\#\#\\
lmc1-106&&&&&&&&&1.73&\#\#\\
lmc1-109&&&&&&&&&2.06&\#\#\\
lmc1-112&&&&&&&&&1.76&\#\#\\
lmc2-16&&&&&&&&&ae&\#\#\\
lmc2-31&&&&&&&&&-6.80&\#\#\\
lmc2-43&&&&&&&&&-26.57&\#\#\\	
lmc3-1&10.96&8.10&0.87&1.91&1.92&0.82&--&1.65&1.59&\\
lmc3-5&5.34&8.32&0.12&6.17&7.39&--&--&7.49&4.56&\\
lmc3-12&2.63&10.55&--&8.08&7.99&--&--&6.64&3.10&\\						
lmc3-20&5.32&	6.13&--&2.25	&2.32&--&--&--&-3.96&\\
lmc3-23	&5.10&	4.78&	-1.17	&2.52	&--&--&--&--&-2.65&\\
lmc3-24	&0.91	&3.73&--&3.43&3.35&--&--&3.00&&\#\\	
lmc3-30	&9.28&	8.05&--&4.92	&3.75&--&--&--&-30.69&\\
lmc3-33	&7.24&	7.69&--&3.36	&3.39&--&--&2.71&&\#\\	
lmc3-37	&2.71&	4.24&--&--&--&--&--&1.98&&\\	
lmc3-39	&--&	2.03&	0.73	&1.39	&2.43&--&0.69&--&&\#\\		
lmc3-44	&--&0.68&--&--&--&--&--&-1.13	&-7.78&\\
lmc3-49&0.42&3.68&1.24&3.17&3.75&--&0.84&1.80&-8.41&\\
\hline
\end{tabular}
\end{table*}
\renewcommand{\thetable}{12}
\begin{table*}
\begin{center}
\caption{List of spectral lines observed in the sample of Be candidate stars (spectra obtained from 2.5m LCO) considered in the present study along with their equivalent widths.}
\end{center}
\begin{tabular}{lrrrrrrrrrrr}
\hline
\hline
star&	H$\delta$&	H$\gamma$&	He{\sc i}&	He{\sc i}&	Fe{\sc ii}&	He{\sc i}&	H$\beta$&	He{\sc i}&	He{\sc i}&SiII&		Fe{\sc ii}\\
&	4101&	4340	&4387&	4471&	4629	&4713&	4861	&4922	&5015&	5052	&5169\\
\hline
lmc1-1&	0.52&	1.23	&--	&1.57	&--	&--	&-1.61	&--	&0.22	&--	&--\\
lmc1-2	&2.61&	3.49	&--	&--	&0.31	&0.14	&3.88	&0.91	&0.29&	0.27&	1.16\\
lmc1-3&	--&	3.16&	-0.68&	1.17&	0.19&	0.6&	2.49	&0.72&	0.6&	--&	--\\
lmc1-4	&0.62&	2.34&	--&	--&	--&	0.07&	2.48&	0.87&	0.23&	0.45	&1.55\\
lmc1-5	&1.57	&2.95	&0.21&	0.27&	--	&0.33&	6.74&0.15&	--&	--	&1.05\\
lmc1-6	&1.52	&5.63&	1.49	&--&	--&	--	&6.73&	--	&--&	--&	0.52\\
lmc1-7	&--&	2.27&	1.43&	0.82	&--&	0.37	&2.4&	0.81	&0.99	&0.53&	0.11\\
lmc1-9	&--	&3.39&	--&	0.45&	--&	--&	3.36&	2.48&	--&	--&	1.4\\
lmc1-10	&3.98&	5.24&	0.45&	0.23&	--&	--&	5.87&	0.15&	0.24&	--&	--\\
lmc1-11	&--&	1.61&	1.05&	0.54&	--&	--&	2.31&	0.61&	--&	--&	--\\
lmc1-12	&--&	1.42	&0.71&	1.15&	--&	0.49&	0.77&	0.67&	0.43&	0.27&	0.11\\
lmc1-15	&--&	10.7	&1.63&	--	&--&	--&	7.78&	--	&--&	--&	--\\
lmc1-114&	5.34&	6.52&	--&	--&	--&	--&	5.95&	--&	--&	--&	--\\
lmc1-115&	--&	5.13&	0.97&	2.64&	--&	1.16&	2.13&	--&	--&	--&	1.96\\
lmc1-116&	--&	1.32&	--&	--&	--&	--&	-0.41&	0.6&	--&	--&	--\\
lmc1-117&	0.55&	0.42&	--&	0.14&	-0.23&	-0.13&	-1.53	&-0.22&	-0.3	&--&	--\\
smc1-40	&2.19	&1.75	&--&	1.12	&1.56&	0.34&	3.69&	1.39&	0.66&	0.39	&-0.26\\
smc1-42	&--&	1.32&	--&	--&	--&	--&	-2.82&	--&	0.88&	0.26&	-0.75\\
smc1-44	&--&	2.76&	0.95&	0.98&--&	--&	0.45&	2.97&	2.11&	-0.35	&--\\
smc1-47	&--&	2.11&	2.75	&1.13&	--	&--	&3.18&	--&	--&	--&	--\\
smc1-50&3.74	&1.16&	1.82	&1.87&	-0.61&	0.13	&4.58	&0.89	&--&	0.45	&-2.27\\
smc1-52	&--&	3.48&	--	&--&	--&	0.27&	2.62&	0.84	&0.66	&--	&-0.11\\
smc1-53	&--&	-6.29	&--&	--&	--&	--&	-8.51&	--&	--&	--&	--\\
smc1-56	&--&	--&	--&	--&	--&	--&	--&	--&	--&	--&	--\\
smc1-57	&--&	0.56&	--&	--&	--&	--&	1.47&	0.61&	--&	--&	--\\
smc1-58&--&	--&	--&	--&	--&	--&	-2.01&   --&	--&	--&	--\\
smc1-59&--&	--&	1.22&	--&	--&	--&	-2.95	&--&	--&	--&	--\\
smc1-60	&--&	-3.36	&--&	--&	--	&--&	-9.27	&--&	--&	--&	--\\
lmc2-1	&--&	1.55&	0.58&	0.69&	0.18&	0.76&	1.7&	0.6&	0.21&	--&	--\\
lmc2-2	&--&	--&	--&	--&	--&	--&	0.58&	0.38&	-1.36&	-0.85&	3.56\\
lmc2-3	&--&	0.97&	0.6	&0.86	&--&	0.13	&-1.95	&0.52	&--&	0.31&	-0.31\\
lmc2-4	&--&	--&	--&	--&	--&	--&	-2.97&	--&	--&	--&	--\\
lmc2-5	&--&	0.59&	--&	--&	--&	--&	-1.89&	0.46&	--&	--&	--\\
lmc2-6	&--&	0.46&	--&	0.32&	--&	--&	--&	0.48&	--&	--&	--\\
lmc2-7	&--&	0.97&	0.27&	0.39&	--&	0.19&	-0.34	&0.36	&--&	--&	--\\
lmc2-8	&--&	-0.66	&--&	--&	--&	--&	-2.72&	--&	-0.62&	--&	--\\
lmc2-9	&--&	1.37	&0.56&	0.88&	--&	0.15&	0.47&	0.85&	0.47	&--	&--\\
lmc3-6&	--&	5.24	&7.19	&7.04	&5.28	&7.10	&7.47	&--	&--	&--	&--\\
smc2-38	&5.25	&8.22&	--&	0.31&	--&	--&	6.52&	0.61&	0.29&	-0.54&	0.83\\
smc2-39	&--&	1.3&	1.32&	1.21&	1.28&	--&	-5.01	&0.63&	--&	--&	-0.33\\
smc2-40	&--&	1.79&	0.86&	0.73	&0.17&	0.51&	2.41&	0.69	&0.22&	--&	-0.29\\
smc2-45	&--&	--&	--	&--&	--&	0.32&	1.48&	1.17&	0.33&	1.23&	--\\
smc3-30	&0.14&	1.47&	0.55&	0.69&	--	&--&	1.81&	1.13	&0.13&	--&	--\\
smc3-31	&4.28&	5.21&	--&	--&	--&	--&	4.78&	0.26&	--&	--&	--\\
smc3-32	&2.96&	3.92&	1.73&	1.84&	--&	1.2&	0.88&	--&	0.75	&--&	--\\
smc3-33	&--&	--&	--&	--	&--&	--&	0.2&	0.21&	-0.34&	--&	--\\
smc3-35	&--&	--&	--&	0.91	&--&	--&	-3.86&	0.37&	--	&0.44&	--\\
\hline
\end{tabular}
\end{table*}
\begin{figure}
\epsfxsize=10truecm
\centerline{\epsffile{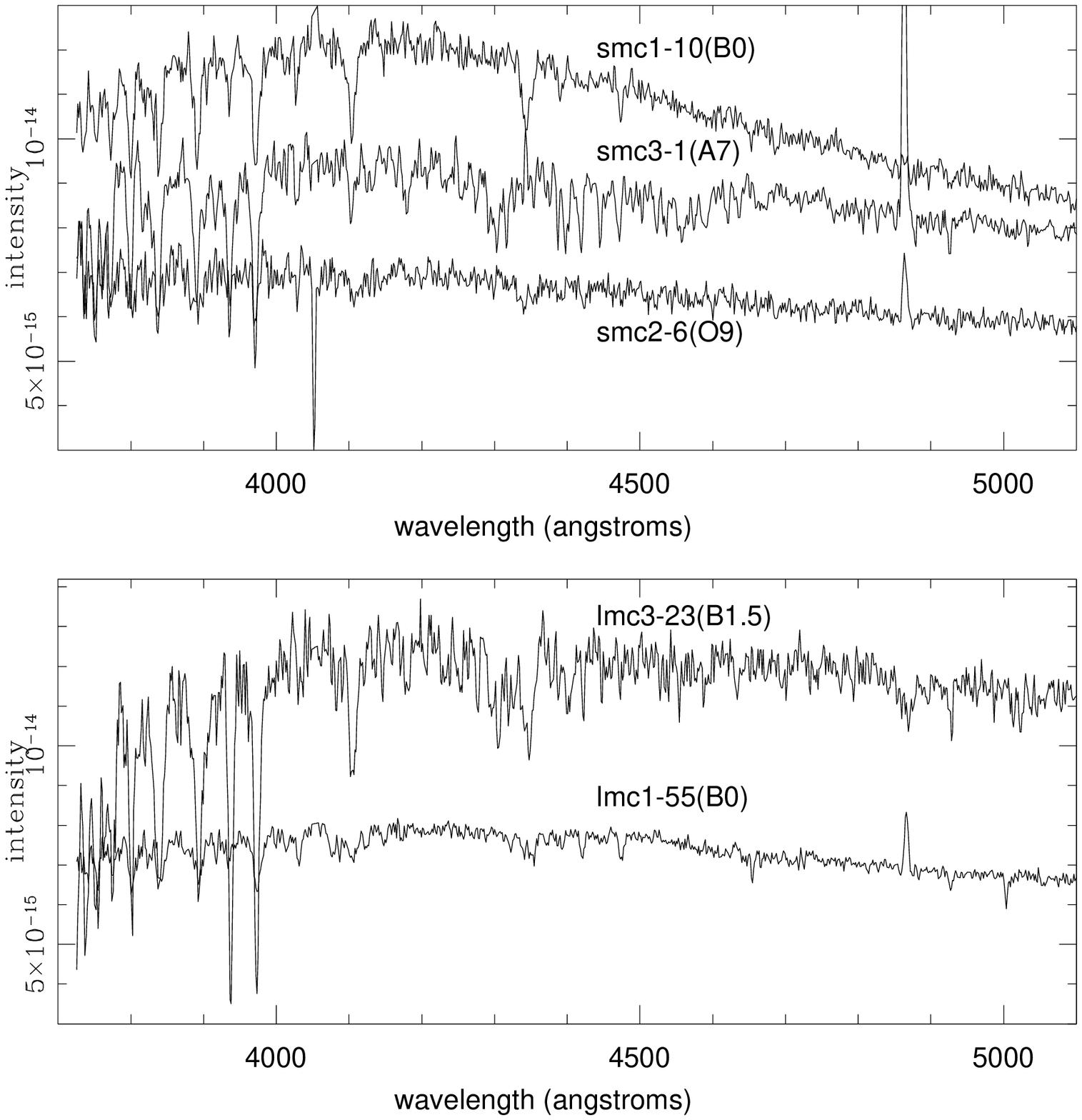}}
\caption{
Representative sample of spectra of Be candidate stars in the LMC and the SMC, in the Blue region
}
\end{figure}


\begin{figure}
\epsfxsize=10truecm
\centerline{\epsffile{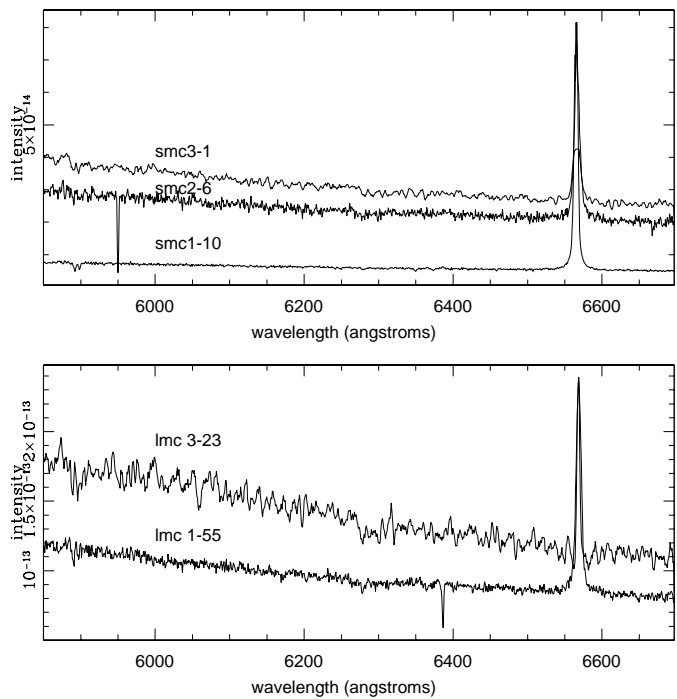}}
\caption{Representative sample of spectra of Be candidate stars in the LMC and the SMC, in the red region
}
\end{figure}
\section{Discussion - comparison between the Clouds}

We studied the NIR properties of Be star candidates in the LMC and the SMC by combining the optical and the NIR magnitudes of a  sample of 1640 stars in the LMC and 839 stars in the SMC. We also used spectra of 109 of the type 1, type 2 and type 3 stars to understand their nature.  In the discussion below, we derive the fraction of various types of stars, which are candidate CBe stars, in the two galaxies studied. The CBe mechanism is expected to be more efficient in the SMC due to its lower metallicity, when compared to the LMC. On the other hand, at low metallicity the kappa-mechanism should be less efficient than at higher metallicity
\citep{salmon2009,miglio2007}. Also, the CBe stars show variability in H$_\alpha$ emission. These factors can affect the ratios discussed below.
 
Within the area surveyed, the LMC and the SMC have varying fractions of various types of stars. The relative fraction of the 4 types were found to be different. In both the galaxies, type 4 is the dominant type, with
more than half of the stars belonging to this group. In the case of the LMC, type 1 stars form one quarter of the population and type 2 and 3 are the least populated group. In the SMC, the second dominant group in the type 2 stars, followed by type 1 and type 3 stars. This trend is also seen if we estimate the ratio of the $LMC/SMC$ for various types. If we take the total sample, the $LMC/SMC$ fraction is 2.3. Now, we can compare this fraction for various types and see how they deviate from the average. These values are tabulated in table 3.
The fraction of stars in the cross-correlated sample are also given in the same table. Both the values are similar suggesting that the cross-correlated sample has similar fractions of the four types of stars.
It can be seen that the type 4 stars have similar fraction (2.14) as the average. Thus, the type 4 stars are similarly populated in the LMC and the SMC. The fraction of the type 3 stars are slightly less than the average, but can  be considered to be similarly populated in the LMC and the SMC. The type 2 stars are found to have a low value of the fraction (0.97), which suggests that these stars are more populated (more than twice) in the SMC, when compared to the LMC. On the other hand, the value of the fraction for type 1 stars
is very high (4.17), when compared to the average. This suggest that the type 1 stars are more populated in the LMC, when compared to the SMC. To summarise, we find that the types 3 and 4 are equally abundant in both the Clouds. Type 1 stars are almost twice abundant in the LMC, whereas type 2 stars are twice abundant in the SMC. 

After combining the optical and the NIR properties, we find that the type 4 stars have a new subgroup of stars with slightly different photometric properties. The new group is found to be about 1/5 of the type 4 population and they could be highly reddened O,B type stars. This new group does not have any NIR excess. This group is not found among the SMC type 4 stars. We have found that the type 4 stars are similarly populated in both the galaxies. Hence if it exists, the new group should have been identified in the SMC as well. Thus it is likely that this group exists only in the LMC. It is important to study this group of stars in detail using spectroscopy to understand their nature. If we remove this subgroup from the type 4 sample, then the type 4 stars are overabundant in the SMC, by about 1.3. 

The spectroscopic analysis suggests that most of the stars studied here show emission features and suggest that they have circumstellar material. Majority of the type 1 and type 2 stars are found to belong to early B type where as type  3 stars were found to belong to A-F type. The spectroscopic analysis supports the photometric result that most of the stars studied here belong to early  type stars, with emission supporting circumstellar material. The type 3 stars are found to belong to a mix of various types of stars and are unlikely to be CBe stars. Our analysis finds that the type 2 stars are likely to be CBe stars. They are unlikely to be pre-MS stars. The interesting fact is that, their number is found to be twice in the SMC when compared to the LMC. Type 1 stars are also found to be similar to the CBe stars.  It is known that due to the lower metallicity of the SMC, the fraction of CBe stars are expected to be more in the SMC. \citet{martayan2010} found that the CBe phenomenon is about $\sim$ 3-5 times enhanced in the SMC, when compared to the Galaxy. \citet{maeder1999} found that the $Be/B$ fraction in the SMC is about 2.4 times that in our Galaxy. Thus, the true CBe stars should be more populated in
the SMC than in the LMC.  The enhancement from the LMC to the SMC is expected to be less than the above range. The type 2 stars follow this trend, with a enhancement of 2.4 and it is quite likely that the type 2 candidates are similar to the Galactic Be stars. This argument is simply based on the metallicity difference and nothing else and we find that the full photometric sample of type 1 does not follow the metallicity trend. If we consider only stars with H$\alpha$ emission among the type 1 stars (which can be confirmed as CBe stars because of emission), we estimate a ratio of (6/22=26\%) in the
LMC and (22/31=71\%) in the SMC.
Therefore, the CBe stars among the type 1 are more in the SMC by 2.6 times. This enhancement is in agreement with \citet{bonanos2010}
and \citet{martayan2010}. The above ratio could be affected by the variability
in H$\alpha$ emission of CBe stars, as seen in LMC1-12. Therefore, the ratio
derived above could be a lower estimate for the LMC, as well as for the SMC.
The fact that only a fraction of the type 1 stars show emission lines in the
spectra suggest that the type 1 could be a heterogeneous group,
consisting of CBe stars and other variables, which do not show H$\alpha$ emission.
In the case of  type 4  stars are similar to CBe stars (without the new subclass in the LMC), we find an enhancement of 1.4 times in the SMC. 
  
 In order to check whether any of the stars studied here are part of star clusters,
we performed a cross-correlation between all the 4 types and the star clusters in both the galaxies.
We used the catalog of \citet{pietrzynski1998,pietrzynski1999} for the SMC and LMC
clusters respectively. We searched for stars located within the radius of
the cluster. We found that about 13\% of the stars, mostly type 4, are probably located in the SMC clusters, and 1\% could belong to the candidate clusters. In the case of the LMC,
22\% of the stars are probably located in the LMC clusters. Thus the majority
of the candidate Be stars in both the galaxies are probably located in the field.
 
This study points to the fact that the photometric variability is a very effective and efficient tool to identify large number of candidate Be stars as well as a variety of binaries in the L\&SMC. \citet{mennickent2002} and \citet{sabogal2005} identified these stars using the OGLEII variability data and followed up with IR photometry and spectral analysis. The next generation OGLE III scans cover much larger area of the LMC and the SMC and the photometric variability data are also available. We suggest that similar studies based on database such as OGLE III will be very useful for comparative studies of CBe candidates as well as binaries in these two galaxies. 
\section {Conclusions}
The conclusions are summarised below:\\

$\bullet$ We combined the optically identified sample of candidate Be stars with the IRSF catalog to obtain the $B, V, J, H$ and $K_s$ magnitudes of all the four types of stars in the LMC (with $\sim$ 70\% cross identification) and the SMC ($\sim$ 80\%). Majority of the sample shows NIR properties similar to the Galactic Be stars. 

$\bullet$ Spectra of 120 stars belonging to the types 1, 2 and 3 were analysed to study their spectral properties. Majority of the stars showed emission lines in the spectra suggesting the presence of circumstellar material and were found to belong to early spectral type.

$\bullet$ The above two results indicate that the photometric variability is a very effective and efficient tool to identify a large number of candidate Be stars and binaries in the L\&SMC.

$\bullet $ We find that type 1, type 2 and type 3 stars have more or less similar spectral and NIR 
properties in the LMC and in the SMC. On the other hand, type 4 stars are found to have a subgroup in the LMC, with different optical and NIR properties. This subgroup is not found in the SMC or in our Galaxy. These stars do not have NIR ecxess, show large reddening, but are not located in regions with high reddening. The reddening corrected magnitudes make them the most brightest and massive stars in the sample. Detailed spectroscopic studies are needed to understand these enigmatic candidates. This new subclass is $\sim$ 18\% of the type 4 sample. The main type 4 sample  is $\sim$ 49\% of the total sample, whereas the SMC has $\sim$ 65\% type 4 stars. 

$\bullet$ Type 3 stars, the least populated type, are found to belong to B and A spectral types. These stars in the LMC have relatively less H$\alpha$ $EW$ when compared to those in the SMC. Some stars are found to have NIR excess and the H$_\alpha$ $EW$ correlates with the $(H-K)_0$ values, suggesting dust in the circumstellar material. These stars also show relatively large reddening. These stars are  possibly a mix of various types of stars like interacting binaries and DPVs, supporting the suggestion by \citet{mennickent2002,mennickent2006}.

$\bullet$ Type 2 stars in the LMC and the SMC show similar spectral and NIR properties. Their NIR properties are similar to the Galactic CBe stars, suggesting that they are likely to be CBe stars. These stars are found in larger fraction in the SMC ($\sim$ 14.5\%), when compared to the LMC ($\sim$ 6\%).

$\bullet$ The type 1 stars are relatively more in the LMC ($\sim$ 24\% ) when compared to the SMC ($\sim$ 13\%). The SMC type 1 stars are early type stars with large H$\alpha$ $EW$ and this class has properties similar to CBe stars. Some of the  type 1 stars in the LMC do not show evidence of emission and their $V_0$ magnitudes suggest that they could belong to late B spectral types. 

$\bullet$ It is known that due to the lower metallicity of the SMC, the fraction of CBe stars is expected to be more in the SMC. Thus, the true CBe stars should be more populated in the SMC than in the LMC. The type 4 stars (without the subclass in the LMC) are more in the SMC by a factor of about 1.4. The type 2 stars follow a similar trend, with an enhancement of 2.4. The spectroscopic
sample of type 1 stars which show H$_\alpha$ in emission and confirmed as CBe stars are more abundant
in the SMC by a factor of 2.6. 

\section{Acknowledgment}
PKT acknowledges the financial assistance extended by Christ University, Bangalore, India.
REM acknowledges support by Fondecyt grant 1070705, 1110347, the Chilean 
Center for Astrophysics FONDAP 15010003 and  from the BASAL
Centro de Astrof\'isica y Tecnologias Afines (CATA) PFB--06/2007. BS acknowledges support from the Faculty of Science,  Universidad de los Andes, Colombia.This research has made use of the IRSF Magellanic Clouds Point Source Catalog, distributed by the National Optical Astronomy Observatories, which are operated by the Association of Universities for Research in Astronomy, Inc., under cooperative agreement with the National Science Foundation.
This publication makes use of data products from the Two Micron All Sky Survey, which is a 
joint project of the University of Massachusetts and the Infrared Processing and Analysis 
Center/California Institute of Technology, funded by the National Aeronautics and Space Administration 
and the National Science Foundation.\\


\begin{thebibliography}{}

\bibitem[\protect\citeauthoryear{{Bessell} \& {Brett}}{{Bessell} \&
  {Brett}}{1988}]{bessell1988}
{Bessell} M.~S.,  {Brett} J.~M.,  1988, \pasp, 100, 1134

\bibitem[\protect\citeauthoryear{{Bonanos}, {Lennon}, {K{\"o}hlinger}, {van
  Loon}, {Massa}, {Sewilo}, {Evans}, {Panagia}, {Babler}, {Block}, {Bracker} \&
  {Engelbracht}}{{Bonanos} et~al.}{2010}]{bonanos2010}
{Bonanos} A.~Z.,  {Lennon} D.~J.,  {K{\"o}hlinger} F.,  {van Loon} J.~T.,
  {Massa} D.~L.,  {Sewilo} M.,  {Evans} C.~J.,  {Panagia} N.,  {Babler} B.~L.,
  {Block} M.,  {Bracker} S.,    {Engelbracht} C.~W.,  2010, \aj, 140, 416

\bibitem[\protect\citeauthoryear{{Carpenter}}{{Carpenter}}{2001}]{carpenter200%
1}
{Carpenter} J.~M.,  2001, \aj, 121, 2851

\bibitem[\protect\citeauthoryear{{de Wit}, {Beaulieu}, {Lamers}, {Lesquoy} \&
  {Marquette}}{{de Wit} et~al.}{2003}]{dewit2003}
{de Wit} W.~J.,  {Beaulieu} J.-P.,  {Lamers} H.~J.~G.~L.~M.,  {Lesquoy} E.,
  {Marquette} J.-B.,  2003, \aap, 410, 199

\bibitem[\protect\citeauthoryear{{Dougherty}, {Waters}, {Burki}, {Cote},
  {Cramer}, {van Kerkwijk} \& {Taylor}}{{Dougherty}
  et~al.}{1994}]{dougherty1994}
{Dougherty} S.~M.,  {Waters} L.~B.~F.~M.,  {Burki} G.,  {Cote} J.,  {Cramer}
  N.,  {van Kerkwijk} M.~H.,    {Taylor} A.~R.,  1994, \aap, 290, 609

\bibitem[\protect\citeauthoryear{{Harris} \& {Zaritsky}}{{Harris} \&
  {Zaritsky}}{2004}]{Harris2004}
{Harris} J.,  {Zaritsky} D.,  2004, \aj, 127, 1531

\bibitem[\protect\citeauthoryear{{Hern{\'a}ndez}, {Calvet}, {Hartmann},
  {Brice{\~n}o}, {Sicilia-Aguilar} \& {Berlind}}{{Hern{\'a}ndez}
  et~al.}{2005}]{hernaandez2005}
{Hern{\'a}ndez} J.,  {Calvet} N.,  {Hartmann} L.,  {Brice{\~n}o} C.,
  {Sicilia-Aguilar} A.,    {Berlind} P.,  2005, \aj, 129, 856

\bibitem[\protect\citeauthoryear{{Indu} \& {Subramaniam}}{{Indu} \&
  {Subramaniam}}{2011}]{indu2011}
{Indu} G.,  {Subramaniam} A.,  2011, \aap, 535, A115

\bibitem[\protect\citeauthoryear{{Jacoby}, {Hunter} \& {Christian}}{{Jacoby}
  et~al.}{1984}]{jacoby1984}
{Jacoby} G.~H.,  {Hunter} D.~A.,    {Christian} C.~A.,  1984, \apjs, 56, 257

\bibitem[\protect\citeauthoryear{{Kato}, {Nagashima}, {Nagayama}, {Kurita},
  {Koerwer}, {Kawai}, {Yamamuro}, {Zenno}, {Nishiyama} \& {Baba}}{{Kato}
  et~al.}{2007}]{kato2007}
{Kato} D.,  {Nagashima} C.,  {Nagayama} T.,  {Kurita} M.,  {Koerwer} J.~F.,
  {Kawai} T.,  {Yamamuro} T.,  {Zenno} T.,  {Nishiyama} S.,    {Baba} D.,
  2007, \pasj, 59, 615

\bibitem[\protect\citeauthoryear{{Keller}, {Wood} \& {Bessell}}{{Keller}
  et~al.}{1999}]{keller1999}
{Keller} S.~C.,  {Wood} P.~R.,    {Bessell} M.~S.,  1999, \aaps, 134, 489

\bibitem[\protect\citeauthoryear{{Koornneef}}{{Koornneef}}{1983}]{koornneef198%
3}
{Koornneef} J.,  1983, \aap, 128, 84

\bibitem[\protect\citeauthoryear{{Maeder}, {Grebel} \& {Mermilliod}}{{Maeder}
  et~al.}{1999}]{maeder1999}
{Maeder} A.,  {Grebel} E.~K.,    {Mermilliod} J.-C.,  1999, \aap, 346, 459

\bibitem[\protect\citeauthoryear{{Martayan}, {Baade} \& {Fabregat}}{{Martayan}
  et~al.}{2010}]{martayan2010}
{Martayan} C.,  {Baade} D.,    {Fabregat} J.,  2010, \aap, 509, A11

\bibitem[\protect\citeauthoryear{{Martayan}, {Fr{\'e}mat}, {Hubert}, {Floquet},
  {Zorec} \& {Neiner}}{{Martayan} et~al.}{2006}]{martayan2006}
{Martayan} C.,  {Fr{\'e}mat} Y.,  {Hubert} A.-M.,  {Floquet} M.,  {Zorec} J.,
   {Neiner} C.,  2006, \aap, 452, 273

\bibitem[\protect\citeauthoryear{{Martayan}, {Fr{\'e}mat}, {Hubert}, {Floquet},
  {Zorec} \& {Neiner}}{{Martayan} et~al.}{2007}]{martayan2007}
{Martayan} C.,  {Fr{\'e}mat} Y.,  {Hubert} A.-M.,  {Floquet} M.,  {Zorec} J.,
   {Neiner} C.,  2007, \aap, 462, 683

\bibitem[\protect\citeauthoryear{{Mathew}, {Subramaniam} \& {Bhatt}}{{Mathew}
  et~al.}{2008}]{mathew2008}
{Mathew} B.,  {Subramaniam} A.,    {Bhatt} B.~C.,  2008, \mnras, 388, 1879

\bibitem[\protect\citeauthoryear{{Mennickent}, {Cidale}, {D{\'{\i}}az},
  {Pietrzy{\'n}ski}, {Gieren} \& {Sabogal}}{{Mennickent}
  et~al.}{2005}]{mennickent2005}
{Mennickent} R.~E.,  {Cidale} L.,  {D{\'{\i}}az} M.,  {Pietrzy{\'n}ski} G.,
  {Gieren} W.,    {Sabogal} B.,  2005, \mnras, 357, 1219

\bibitem[\protect\citeauthoryear{{Mennickent}, {Cidale}, {Pietrzy{\'n}ski},
  {Gieren} \& {Sabogal}}{{Mennickent} et~al.}{2006}]{mennickent2006}
{Mennickent} R.~E.,  {Cidale} L.,  {Pietrzy{\'n}ski} G.,  {Gieren} W.,
  {Sabogal} B.,  2006, \aap, 457, 949

\bibitem[\protect\citeauthoryear{{Mennickent}, {Pietrzy{\'n}ski}, {Gieren} \&
  {Szewczyk}}{{Mennickent} et~al.}{2002}]{mennickent2002}
{Mennickent} R.~E.,  {Pietrzy{\'n}ski} G.,  {Gieren} W.,    {Szewczyk} O.,
  2002, \aap, 393, 887

\bibitem[\protect\citeauthoryear{{Mennickent} \& {Smith}}{{Mennickent} \&
  {Smith}}{2010}]{mennickent2010}
{Mennickent} R.~E.,  {Smith} M.~A.,  2010, \mnras, 407, 734

\bibitem[\protect\citeauthoryear{{Meyer}, {Calvet} \& {Hillenbrand}}{{Meyer}
  et~al.}{1997}]{meyer1997}
{Meyer} M.~R.,  {Calvet} N.,    {Hillenbrand} L.~A.,  1997, \aj, 114, 288

\bibitem[\protect\citeauthoryear{{Miglio}, {Montalb{\'a}n} \&
  {Dupret}}{{Miglio} et~al.}{2007}]{miglio2007}
{Miglio} A.,  {Montalb{\'a}n} J.,    {Dupret} M.-A.,  2007, \mnras, 375, L21

\bibitem[\protect\citeauthoryear{{Pietrzynski}, {Udalski}, {Kubiak},
  {Szymanski}, {Wozniak} \& {Zebrun}}{{Pietrzynski}
  et~al.}{1998}]{pietrzynski1998}
{Pietrzynski} G.,  {Udalski} A.,  {Kubiak} M.,  {Szymanski} M.,  {Wozniak} P.,
    {Zebrun} K.,  1998, \actaa, 48, 175

\bibitem[\protect\citeauthoryear{{Pietrzynski}, {Udalski}, {Kubiak},
  {Szymanski}, {Wozniak} \& {Zebrun}}{{Pietrzynski}
  et~al.}{1999}]{pietrzynski1999}
{Pietrzynski} G.,  {Udalski} A.,  {Kubiak} M.,  {Szymanski} M.,  {Wozniak} P.,
    {Zebrun} K.,  1999, \actaa, 49, 521

\bibitem[\protect\citeauthoryear{Sabogal et al.} {2012}]{sabogal2012}Sabogal B.E., et al. 2012, in preparation

\bibitem[\protect\citeauthoryear{{Sabogal}, {Garc{\'{\i}}a-Varela} \&
  {Mennickent}}{{Sabogal} et~al.}{2011}]{sabogal2011}
{Sabogal} B.,  {Garc{\'{\i}}a-Varela} A.,    {Mennickent} R.~E.,  2011, in
  {C.~Neiner, G.~Wade, G.~Meynet, \& G.~Peters} ed., IAU Symposium Vol.~272 of
  IAU Symposium, {A spectroscopic study of Be-like variable stars in the Small
  Magellanic Cloud}.
pp 308--309

\bibitem[\protect\citeauthoryear{{Sabogal}, {Mennickent}, {Pietrzy{\'n}ski} \&
  {Gieren}}{{Sabogal} et~al.}{2005}]{sabogal2005}
{Sabogal} B.~E.,  {Mennickent} R.~E.,  {Pietrzy{\'n}ski} G.,    {Gieren} W.,
  2005, \mnras, 361, 1055

\bibitem[\protect\citeauthoryear{{Salmon}, {Montalb{\'a}n}, {Miglio}, {Morel},
  {Dupret} \& {Noels}}{{Salmon} et~al.}{2009}]{salmon2009}
{Salmon} S.,  {Montalb{\'a}n} J.,  {Miglio} A.,  {Morel} T.,  {Dupret} M.-A.,
   {Noels} A.,  2009, in {J.~A.~Guzik \& P.~A.~Bradley} ed., American Institute
  of Physics Conference Series Vol.~1170 of American Institute of Physics
  Conference Series, {The Enigma of B-type Pulsators in the SMC}.
pp 385--387

\bibitem[\protect\citeauthoryear{{Udalski}, {Szymanski}, {Kubiak},
  {Pietrzynski}, {Soszynski}, {Wozniak} \& {Zebrun}}{{Udalski}
  et~al.}{2000}]{udalski2000}
{Udalski} A.,  {Szymanski} M.,  {Kubiak} M.,  {Pietrzynski} G.,  {Soszynski}
  I.,  {Wozniak} P.,    {Zebrun} K.,  2000, \actaa, 50, 307

\bibitem[\protect\citeauthoryear{{Udalski}, {Szymanski}, {Kubiak},
  {Pietrzynski}, {Wozniak} \& {Zebrun}}{{Udalski} et~al.}{1998}]{udalski1998}
{Udalski} A.,  {Szymanski} M.,  {Kubiak} M.,  {Pietrzynski} G.,  {Wozniak} P.,
    {Zebrun} K.,  1998, \actaa, 48, 147

\end{thebibliography}

\appendix{Appendix}

\begin{center}
\scriptsize
\renewcommand{\thetable}{1}
\begin{longtable}{lllllll}

\caption[Log of spectroscopic observations from 1.5m CTIO Telescope ]{Log of spectroscopic observations from 1.5m CTIO Telescope }
\\
\hline \multicolumn{1}{l}{\textbf{Star}} & \multicolumn{1}{l}{\textbf{OGLE ID}} & \multicolumn{1}{l}{\textbf{Blue spectra}} & \multicolumn{1}{l}{\textbf{Exp.Time}} & \multicolumn{1}{l}{\textbf{Red Spectra} }& \multicolumn{1}{l}{\textbf{Exp time} }& \multicolumn{1}{l}{\textbf{Comment on H$\alpha$ profile}} \\

 &  & \multicolumn{1}{l}{\textbf{Date of OBS}} &  & \multicolumn{1}{l}{\textbf{Date of OBS} }& & \\
\hline  
\endfirsthead
 \multicolumn{7}{l}%
 {{\bfseries \tablename\ \thetable{} -- continued from previous page}} \\
\hline\multicolumn{1}{l}{\textbf{Star Type}} &
\multicolumn{1}{l}{\textbf{OGLE ID}} &
 \multicolumn{1}{l}{\textbf{Blue spectra}} &
\multicolumn{1}{l}{\textbf{Exp.Time}} &
\multicolumn{1}{l}{\textbf{Red Spectra}} &
\multicolumn{1}{l}{\textbf{Exp. time}} & 
\multicolumn{1}{l}{\textbf{Comment on H$\alpha$ profile}} \\ 

&  & \multicolumn{1}{l}{\textbf{Date of OBS}} &  & \multicolumn{1}{l}{\textbf{Date of OBS} }& & \\
\hline
\endhead

\hline \multicolumn{7}{r}{{continued on next page}} \\ \hline
\endfoot

\hline \hline
\endlastfoot
smc1-2&OGLE003918.20-733656.6&10-10-2002&900&--&--&--\\
smc1-7&OGLE005059.21-724357.3&10-10-2002&900&12-10-2002&900&strong emission\\
smc1-10&OGLE005224.40-724038.6&09-10-2002&900&12-10-2002&900&weak emission\\
smc1-11&OGLE005227.51-732001.2&09-10-2002&900&--&--&--\\
smc1-16&OGLE005535.60-731029.2&--&--&11-10-2002&900&weak emission\\
smc1-17&OGLE005504.55-724637.3&09-10-2002&900&--&--&--\\
smc1-18&OGLE005614.63-723755.1&10-10-2002&900&12-10-2002&900&strong emission\\
smc1-19&OGLE005641.86-724425.4&10-10-2002&900&--&--&--\\
smc1-21&OGLE006916.06-722100.3&10-10-2002&900&12-10-2002&900&strong emission\\
smc1-23&OGLE010041.93-723028.6&10-10-2002&900&--&--&--\\
smc1-24&OGLE010043.94-722604.8&10-10-2002&900&--&--&--\\
smc1-25&OGLE010056.79-721635.2&--&--&11-10-2002&900&strong double emission\\
smc1-26&OGLE010058.69-723049.9&09-10-2002&900&11-10-2002&900&weak emission\\
smc1-28&OGLE010213.80-722213.0&09-10-2002&900&11-10-2002&900&strong emission\\
smc1-31&OGLE010542.57-722747.3&10-10-2002&900&--&--&--\\
smc1-36&OGLE010807.38-721932.6&09-10-2002&900&11-10-2002&900&weak emission\\
smc1-37&OGLE010809.52-721932.6&09-10-2002&900&12-10-2002&900&weak emission\\
smc1-38&OGLE010825.82-722327.2&09-10-2002&900&11-10-2002&900&strong emission\\
smc1-39&OGLE005235.60-723751.7&09-10-2002&900&11-10-2002&900&weak emission\\
smc2-4&OGLE004721.86-730650.1&09-10-2002&900&11-10-2002&900&strong emission\\
smc2-6&OGLE004848.83-730620.1&09-10-2002&900&11-10-2002&900&strong emission\\
smc2-8&OGLE004849.49-724800.0&10-10-2002&900&12-10-2002&900&emission in absorption\\
smc2-9&OGLE004938.01-730610.0&09-10-2002&900&11-10-2002&900&weak emission\\
smc2-12&OGLE005224.40-724038.6&09-10-2002&900&--&--&--\\
smc2-14&OGLE005251.97-723508.5&10-10-2002&900&12-10-2002&900& emission\\
smc2-19&OGLE005614.45-724053.2&09-10-2002&900&--&--&--\\
smc2-20&OGLE005618.51-722645.2&09-10-2002&900&--&--&--\\
smc2-23&OGLE005806.05-723544.5&10-10-2002&900&--&--&--\\
smc2-27&OGLE010025.10-724632.2&10-10-2002&900&12-10-2002&900&--\\
smc2-29 &OGLE010401.11-723311.1&09-10-2002&900&11-10-2002&900&strong emission\\
smc2-32&OGLE010447.25-722559.4&10-10-2002&900&12-10-2002&900&emission\\
smc3-1&OGLE004336.91-732637.7&09-10-2002&900&11-10-2002&900& strong double emission\\
smc3-2&OGLE004454.66-732802.9&08-10-2002&600&11-10-2002&300&weak emission\\
smc3-3&OGLE004554.14-731404.3&09-10-2002&900&12-10-2002&300&--\\
smc3-4&OGLE004721.86-730650.1&--&--&11-10-2002&900&strong double emission\\
smc3-5&OGLE004750.14-731316.4&09-10-2002&900&--&--&--\\
smc3-6&OGLE004921.41-725844.9&09-10-2002&900&--&--&--\\
smc3-7&OGLE005025.64-725807.1&09-10-2002&900&--&--&--\\
smc3-8&OGLE005043.44-732705.3&09-10-2009&900&11-10-2002&300&strong emission\\
smc3-9&OGLE005100.18-725303.9&09-10-2002&900&11-10-2002&900&weak emission\\
smc3-10&OGLE005107.59-732636.6&--&--&11-10-2002&900&--\\
smc3-12&OGLE005359.22-723508.9&09-10-2002&900&--&--&--\\
smc3-14&OGLE005520.27-723710.1&10-10-2002&900&12-10-2002&900&weak emission\\
smc3-15&OGLE005745.25-723532.1&09-10-2002&900&--&--&--\\
smc3-16&OGLE005812.58-723048.5&09-10-2002&900&11-10-2002&900&weak emission\\
smc3-17&OGLE005822.04-725522.9&--&--&11-10-2009&900&weak emission\\
smc3-18&OGLE010000.78-725522.9&09-10-2002&900&--&--&weak emission\\
smc3-20&OGLE010325.11-724646.9&09-10-2002&900&11-10-2002&900&strong emission\\
smc3-21&OGLE010452.99-715918.8&09-10-2002&900&11-10-2002&900&weak emission\\
lmc1-8&OGLE050220.61-690240.9&--&--&12-10-2002&900&--\\
lmc1-12&OGLE050351.57-685300.9&--&--&12-10-2002&900&emission\\
lmc1-52&OGLE051651.98-691129.5&--&--&12-10-2002&900&emission\\
lmc1-55&OGLE051747.85-690908.6&10-10-2002&900&11-10-2002&900&emission\\
lmc1-60&OGLE051813.34-691305.4&--&--&12-10-2002&900&--\\
lmc1-106&OGLE053723.19-701349.4&--&--&12-10-2002&900&--\\
lmc1-109&OGLE054259.85-704153.2&--&--&12-10-2002&900&--\\
lmc1-112&OGLE054715.07-705300.6&--&--&12-10-2002&900&--\\
lmc2-16&OGLE050918.50-684225.2&--&--&12-10-2002&900&--\\
lmc2-31&OGLE051824.53-691552.5&--&--&12-10-2002&900&emission double peak\\
lmc2-43&OGLE052455.30-700706.3&--&--&12-10-2002&900&emission\\
lmc3-1&OGLE050052.36-685803.7&09-10-2002&900&11-10-2002&900&--\\
lmc3-5&OGLE050303.70-690615.0&09-10-2002&900&11-10-2002&900&--\\
lmc3-12&OGLE050848.63-684315.6&10-10-2002&900&11-10-2002&900&--\\
lmc3-20&OGLE051418.21-691235.0&10-10-2002&900&11-10-2002&900&weak emission\\
lmc3-23&OGLE051647.54-694415.2&10-10-2002&900&11-10-2002&900&weak emission\\
lmc3-24&OGLE051744.42-692033.3&10-10-2002&900&--&--&--\\
lmc3-30&OGLE051947.82-693912.3&10-10-2002&900&11-10-2002&900&strong double emission\\
lmc3-33&OGLE052032.26-694224.2&10-10-2002&900&--&--&--\\
lmc3-37&OGLE052258.47-692621.0&10-10-2002&900&11-10-2002&900&--\\
lmc3-39&OGLE052402.01-694920.5&10-10-2002&900&--&--&--\\
lmc3-44&OGLE052652.49-693317.2&10-10-2002&900&11-10-2002&900&weak emission\\
lmc3-49&OGLE052938.98-693448.0&10-10-2002&900&11-10-2002&900&weak emission\\
\end{longtable}
\end{center}

\renewcommand{\thetable}{2}
\begin{table*}
\begin{center}
\caption{Log of spectroscopic observations from 2.5m  LCO Telescope }
\end{center}
\scriptsize
\begin{tabular}{llllll}

\hline
\hline
  Star & OGLE ID&Date of Obs& Exp.Time &No. of Exposures&Comment on H$\beta$ profile   \\
\hline
lmc1-1&OGLE050122.35-690004.9&12-11-2003&300&4& emission\\
lmc1-2&OGLE050131.42-684251.0&12-11-2003&300&4\\
lmc1-3&OGLE050132.69-692845.8&13-11-2003&300&3\\
lmc1-4&OGLE050147.90-692424.1&12-11-2003&300&4\\
lmc1-5&OGLE050151.04-692109.6&12-11-2003&300&4\\
lmc1-6&OGLE050158.32-684315.5&13-11-2003&300&5\\
lmc1-7&OGLE050221.24-690451.6&12-11-2003&300&5\\
lmc1-9&OGLE050127.92-685947.7&13-11-2003&300&3\\
lmc1-10&OGLE050234.22-683939.8&13-11-2003&300&4\\
lmc1-11&OGLE050320.42-685315.7&13-11-2003&300&5\\
lmc1-12&OGLE050438.96-691624.6&13-11-2003&300&3\\
lmc1-13&OGLE050426.31-690714.4&13-11-2003&300&5\\
lmc1-15&OGLE050447.46-690552.3&13-11-2003&300&5\\
lmc1-114&OGLE050151.61-690152.8&14-11-2003&300&6\\
lmc1-115&OGLE050428.01-690927.5&14-11-2003&300&6\\
lmc1-116&OGLE050521.34-691550.1&14-11-2003&300&5&emission\\
lmc1-117&OGLE053319.88-701213.1&14-11-2003&300&4\\
lmc2-1&OGLE050017.58-692749.9&15-11-2003&300&4\\
lmc2-2&OGLE050039.16-692003.9&15-11-2003&300&5\\
lmc2-3&OGLE050053.02-692011.5&13-11-2003&300&3&emission\\
lmc2-4&OGLE050103.77-691746.7&14-11-2003&300&4&emission\\
lmc2-5&OGLE050116.74-692027.7&14-11-2003&300&4&emission\\
lmc2-6&OGLE050306.60-691807.5&14-11-2003&300&5\\
lmc2-7&OGLE050340.10-691530.9&15-11-2003&300&3&emission\\
lmc2-8&OGLE050350.56-690223.0&15-11-2003&300&4&emission\\
lmc2-9&OGLE050412.28-685021.7&15-11-2003&300&4\\
lmc3-6&OGLE050343.42-685947.7&13-11-2003&300&4\\
smc1-40&OGLE003631.42-733917.8&11-11-2003&300&5\\
smc1-42&OGLE004045.97-732925.6&11-11-2003&300&4&emission\\
smc1-44&OGLE004225.27-731718.2&11-11-2003&300&5\\
smc1-47&OGLE004626.57-731929.3&13-11-2003&300&5\\
smc1-50&OGLE004701.71-731041.9&12-11-2003&300&5\\
smc1-52&OGLE004801.80-731057.0&13-11-2003&300&5\\
smc1-53&OGLE004803.29-730722.2&14-11-2003&300&6&emission\\
smc1-56&OGLE004916.20-724941.2&14-11-2003&300&5\\
smc1-57&OGLE004930.81-731236.5&14-11-2003&300&5\\
smc1-58&OGLE004948.88-722330.0&14-11-2003&300&4&emission\\
smc1-59&OGLE004958.39-725750.8&14-11-2003&300&5&emission\\
smc1-60&OGLE005018.69-725524.3&14-11-2003&300&5&strong emission\\
smc2-38&OGLE004236.69-733033.1&11-11-2003&300&3\\
smc2-39&OGLE004304.50-730206.3&12-11-2003&300&5&emission\\
smc2-40&OGLE004327.71-731653.9&12-11-2003&300&6\\
smc2-45&OGLE004521.57-731717.7&12-11-2003&300&5\\
smc3-30&OGLE005710.71-722550.2&12-11-2003&300&5\\
smc3-31&OGLE005617.49-730005.1&12-11-2003&300&5\\
smc3-32&OGLE004652.17-731409.2&13-11-2003&300&5\\
smc3-33&OGLE005059.66-725648.3&13-11-2003&300&5\\
smc3-35&OGLE005217.74-725627.9&13-11-2003&300&5&emission\\

\hline

\end{tabular}
\end{table*}
\end{document}